\documentclass[twocolumn,epjc3,final]{svjour3}
\pdfoutput=1

\usepackage[utf8]{inputenc}
\usepackage[T1]{fontenc}

\usepackage[fleqn]{amsmath}
\usepackage{amsfonts}
\usepackage{amssymb}

\usepackage{mathptmx}

\usepackage{siunitx}
	\DeclareSIUnit{\gauss}{G}
	\DeclareSIUnit{\parsec}{pc}

\usepackage{physics}

\usepackage{tensor}

\usepackage{graphicx}
\usepackage{caption}

\usepackage{subfig}
\usepackage{float}
\usepackage{afterpage}

\usepackage{hyperref}
	\hypersetup{%
		pdfauthor={Federico Di Gioia and Giovanni Montani},%
		pdftitle={Linear perturbations of an anisotropic Bianchi~I model with a uniform magnetic field}%
	}
\usepackage{hypernat}
\newcommand{\nepernumber}{\mathrm{e}}
\newcommand{\iunit}{\mathrm{i}}

\DeclareMathOperator{\D}{D}
\DeclareMathOperator{\const}{const}
\DeclareMathOperator{\diag}{diag}

\newcommand{\back}{\mathrm{B}}

\newcommand{\hubble}[1]{\frac{\dot{#1}}{#1}}
\newcommand{\ddotaovera}[1]{\frac{\ddot{#1}}{#1}}
\newcommand{\doubleCoefficient}[2]{{\left({}^{#1}{}_{#2}\right)}}

\journalname{Eur. Phys. J. C}

\begin{document}
\title{Linear perturbations of an anisotropic Bianchi~I model with a uniform magnetic field}

\author{Federico Di Gioia\thanksref{e1,addr1,addr2} \and Giovanni Montani\thanksref{e2,addr1,addr3}}

\thankstext{e1}{e-mail: federico.digioia@uniroma1.it}
\thankstext{e2}{e-mail: giovanni.montani@enea.it}

\institute{Physics Department, ``La Sapienza'' University of Rome P.le A. Moro 5, 00185 Roma, Italy \label{addr1}
           \and
           INFN, Roma 1 Section P.le A. Moro 5, 00185 Roma, Italy \label{addr2}
           \and
           ENEA, C.R. Frascati (Rome), Italy Via E.\ Fermi 45, 00044 Frascati (Roma), Italy \label{addr3}
}

\maketitle

\begin{abstract}
	In this work, we study the effect of a magnetic field on the growth of cosmological perturbations. We develop a mathematical consistent treatment in which a perfect fluid and a uniform magnetic field evolve together in a Bianchi~I universe. We then study the energy density perturbations on this background with particular emphasis on the effect of the background magnetic field. We develop a full relativistic solution which refines previous analysis in the relativistic limit, recovers the known ones in the Newtonian treatment with adiabatic sound speed, and it adds anisotropic effects to the relativistic ones for perturbations with wavelength within the Hubble horizon. This represents a refined approach on the perturbation theory of an isotropic universe in GR, since most of the present studies deal with fully isotropic systems.
\end{abstract}

\section{Introduction\label{sec:intro}}
	The formation of large scale structures across the Universe is one of the most fascinating and puzzling questions, still opened in theoretical cosmology. Among the long standing problems of this investigation area is the determination of the basic nature and dynamics of the cold dark matter~\cite{bib:armendariz-picon-2014}, responsible for the gravitational skeleton on which the baryonic matter falls in, forming the radiative component of the present structures.
	
	However, also the peculiarity of the matter distribution across the Universe, in particular the possibility for large scale filaments~\cite{bib:gheller-2016}, as well as hypotheses for structure fractal dimension~\cite{bib:dickau-2009,bib:grujic-2009} call attention for a deeper comprehension.
	
	In this respect, we observe that the Universe plasma nature, both before the Hydrogen recombination and, for a part in $ 10^5 $ also in the later matter dominated era~\cite{bib:banerjee-2004,bib:lattanzi-2012}, has to be taken into account.
	
 	At the recombination the Universe Debye length is of the order of \SI{10}{\centi\metre} and therefore the implementation of a fluid theory, like General Relativistic Magneto-hydrodynamics is to be regarded as a valid and viable approach to treat the influence of the primordial magnetic field~\cite{bib:giovannini-2004} on the evolution of perturbations~\cite{bib:lattanzi-2012}. Nonetheless, the smallness of such magnetic field, as constrained by the Cosmic Microwave Background Radiation (CMBR) up to \SI{e-9}{\gauss}~\cite{bib:kosowsky-1996,bib:barrow-1997-constraints,bib:barrow-1997,bib:komatsu-2011,bib:paoletti-2011,bib:pogosian-2014,bib:planck-2015}, significantly limits the impact of the plasma nature of the cosmological fluid on the evolution of perturbations. As shown in~\cite{bib:lattanzi-2012,bib:montani-2017}, the presence of the magnetic field is able to trigger anisotropy in the linear perturbations growth and it can be inferred that in the full non-linear\linebreak regime, such anisotropy grows up to account for the formation of large scale filaments.
	
	Apparently, a weak point in the perspective traced above consists of the small plasma component surviving when the Hydrogen recombines and in the observation that the most relevant cosmological scales enter the non-linear regime in such a neutral Universe. Instead, it can be surprisingly\linebreak demonstrated~\cite{bib:banerjee-2004,bib:lattanzi-2012,bib:montani-2017} that the coupling between the neutral and ionized matter is very strong at spatial scale of cosmological interest (for overdensities of mass greater than $ 10^6 $ solar masses, the Ambipolar Reynold number is much greater than unity for redshift $ 10 < z < 1000 $). Thus, the dynamical features, for instance anisotropy, that we recover for the plasma component clearly concern the Universe baryonic component too. This statement is not affected by the presence of dark matter gravitational skeleton in formation, simply because the radiation pressure prevents, up to $ z \sim 100 $ the real fall down of the baryonic fluid into the gravitational well. In fact, the large photon to baryon ratio, about $ 10^9 $ (also constant during the Universe evolution), maintains active a strong Thomson scattering process, even after the hydrogen is recombined into atoms~\cite{bib:weinberg-gravitationAndCosmology,bib:kolb-turner,bib:weinberg-cosmology,bib:banerjee-2004,bib:lattanzi-2012}.
	
	These considerations are to underline that a single fluid General Relativistic Magneto-hydrodynamics formulation is an appropriate tool to investigate the impact of the Universe plasma features on structure formation, at least for a large range of the cosmological thermal history.
	
	In this context many works have been developed, mainly assuming as negligible the backreaction of the magnetic field on the isotropic Universe, see~\cite{bib:barrow-2007} and references therein. However, the presence of a magnetic field rigorously violates the isotropy of the space and the (essentially) flat\linebreak Robertson-Walker geometry must be replaced by a Bianchi I model. This paper faces the general question of how the linear perturbations evolve on a background Bianchi I cosmology, thought as a weak perturbation of the isotropic case, but treated in its full generality for arbitrary large magnetic fields.
	
	We discuss in detail the structure of the perturbation\linebreak equations in the synchronous gauge and the specific form of the spectrum time dependence in specific important limits, like the large scale limit, when the dependence on the wavenumber can be suppressed, and the sub horizon limit, when the dependence on the wavenumber is dominant.
	
	Furthermore, the change of the Jeans scale, when passing from the ionized to the (essentially) recombined Universe, is determined for the small scales, shedding light on the role of the magnetic field and on the real nature of the gauge perturbations.
	
	We recover the slowing-down of the growing mode in super-horizon scales, long known in FRW models. This effect is very small given the upper limits on the cosmological magnetic fields, of order $ \order{v_A^2} \ll 1 $. At sub-horizon scales, we generalise the solutions of~\cite{bib:vasileiou-2015} and~\cite{bib:tseneklidou-2018}, which in turn generalise the results of~\cite{bib:barrow-2007}. While they consider random (i.e.\ isotropic) magnetic fields to preserve the FRW model, we work in the anisotropic case and also consider a nonvanishing sound speed.
	
	Finally, we stress that, along the whole analysis, we compare our results with previous achievements in literature, providing a significant contribution to the understanding of the different effects that the Universe anisotropy, due to the magnetic field, induces on the perturbation evolution and stability.
	
	We notice that there is another paper about this matter~\cite{bib:tsagas-2000}, which was the first analytical study to address this issue. There, the authors study the model in 3 different physical limits with specific anisotropies, while we completely relate the background anisotropy to the magnetic field.
	
	The paper is structured as follows: in section~\ref{sec:basicEquations} we summarize the exact GRMHD equations in the 3+1 covariant formalism; in section~\ref{sec:background} we find the solution for the background Bianchi~I model, then we write the equations for the perturbations in synchronous gauge in section~\ref{sec:perturbedEquations} and we find the gauge modes in section~\ref{sec:gaugeModes}; finally we solve our system in some specific cases in section~\ref{sec:solutionProcedure} and we compare our results with present literature.
	
\section{General properties of the Bianchi~I models}
	As we already said, it is impossible to accommodate a magnetic field in a isotropic model. Moreover, although present observations show that the isotropic FRW model describes very well the present universe, it is only a very special description of the universe towards the initial singularity, while the general one should incorporate anisotropy~\cite{bib:belinskii-1982,bib:landau-classicalFields}.
	
	In the first stage of the universe evolution the matter contribution is negligible, while it is necessary to have a isotropic matter field to achieve the isotropization of the model~\cite{bib:zeldovichNovikov-relativisticAstrophysics2TheStructureAndEvolutionOfTheUniverse,bib:montani-primordialCosmology}. The general solution is constructed through the Bianchi~VIII and~IX models~\cite{bib:belinskii-1982,bib:landau-classicalFields,bib:montani-primordialCosmology}, but we will focus for simplicity on a single Kasner era and so we will use a Bianchi~I model.
	
	The Bianchi~I model is similar to the FRW one, but with three different scale factors. It is intrinsically anisotropic in vacuum, i.e.\ the three cosmic scale factors are never all equal; moreover, in vacuum one of the three scale factor always decreases with time, meaning that one of the spatial direction is contracting.
	
	Near enough to the cosmological singularity, any matter source in the form of perfect fluid energy density, having equation of state $ {p = w \rho} $ always behaves as a test fluid, i.e.\ it induces negligible backreaction, as far a $ {0 < w < 1} $. Since the background magnetic field energy density is a radiation-like term in the Universe and it is associated to an equation of state $ {p = \rho /3} $, near enough to the singularity, we can expect a typical vacuum solution of the Kasner form~\cite{bib:landau-classicalFields,bib:montani-primordialCosmology}.
	
	The more general Bianchi~IX model can be described as a succession of Kasner epochs, in which the different directions exchange time evolutions, alternating moments of growing and decreasing~\cite{bib:montani-primordialCosmology}. For more detailed informations regarding the Bianchi models we recommend~\cite{bib:ryanShepley-homogeneousRelativisticCosmologies}.
	
	Clearly, as soon as the Universe expands enough, the matter source can no longer be negligible and, if the\linebreak pressure term is isotropic, the solution must correspondingly isotropize, i.e.\ the three scale factors tend to be equivalent. This process of isotropization is particularly efficient in the case of an inflationary paradigm~\cite{bib:kirillov-2002,bib:montani-primordialCosmology}, when a vacuum energy, having an equation of state $ {p = -\rho} $ is dominating the Universe dynamics.
	
	The relevance of our study for the structure formation takes place when the isotropization process reduced the\linebreak Bianchi~I cosmology to a flat Robertson-Walker Universe, except for the residual intrinsic anisotropy due to the presence of a background magnetic field.
	
	There exist already a large number of studies regarding Bianchi~I models, analysing cases with different values for the barotropic index~$ {w} $ of the matter source in addition to the magnetic field. \cite{bib:leblanc-1997}~was probably the first to address their stability. \cite{bib:zeldovichNovikov-relativisticAstrophysics2TheStructureAndEvolutionOfTheUniverse}~studies the effect of a pure magnetic matter component, \cite{bib:thorne-1967}~contains analytic solutions for dust~$ {w = 0} $ and radiation~$ {w = 1/3} $, \cite{bib:jacobs-1969}~contains solutions for~$ {w = 1} $ and~$ {1/3 \leq w \leq 1} $ and for the pure magnetic case, \cite{bib:king-2007}~analyses the case of vacuum energy~$ {w = -1} $. The nature of the solutions depends on the values of various constants, it can collapse isotropically or anisotropically, only in the longitudinal or in the transverse direction towards the Big Bang. In general the magnetic fields accelerates expansion (or decelerates collapse) in the transverse direction of the magnetic pressure and it decelerates expansion (or accelerates collapse) in the direction of the magnetic tension. For general properties of the solutions, see~\cite{bib:stephaniKramer-exactSolutionsOfEinsteinsFieldEquations}.
	
	Some interesting cases are analysed in~\cite{bib:thorne-1967}: if~$ {B^2/\rho \rightarrow 0} $ towards the singularity then the magnetic field effects are negligible; if~$ {B^2/\rho} $ does not approach~$ 0 $, then it is constant and both fluids determine the dynamics, or the magnetic field causes a rapid expansion in the transverse direction and this change of the dynamics causes~$ {B^2/\rho \rightarrow 0} $. Moreover, \cite{bib:king-2007}~shows that in presence of a cosmological constant the magnetic field has a strong effect at early times, decelerating the collapse in the transverse direction and accelerating it in the longitudinal one, and is negligible at later times, when the vacuum energy causes accelerated expansion in both directions; the authors also describe the shape of the singularity.
	
	It should be noted that in general the presence of the magnetic field causes a slowing down in the process of\linebreak isotropization, making the shear more important; this way the CMB gives a strong constraint on primordial homogeneous magnetic fields~\cite{bib:barrow-1997-constraints,bib:barrow-1997}.

\section{Basic equations\label{sec:basicEquations}}
	We will now recap the fundamental equations we'll need later; their derivation can be found in~\cite{bib:barrow-2007}. Following~\cite{bib:barrow-2007} we define the magnetic field as the spatial part of the Faraday tensor~$ F_{\mu\nu} $ in the frame comoving with the cosmological fluid; we will use the ideal MHD approximation to turn off the electric field. These equations can be easily obtained in the covariant~3+1 formalism~\cite{bib:ehlers-1961,bib:ellis-1971,bib:ellis-1973,bib:ellis-1998}, as done in~\cite{bib:tsagas-1997,bib:tsagas-2005,bib:barrow-2007,bib:tsagas-2008}; we will solve them, however, in a fixed\linebreak synchronous gauge. We will assume geometric units for the speed of light~$ c $ and Newton's gravitational constant~$ G $ in witch $ {c = 8 \pi G / c^4 = 1} $.
	
	We describe an anisotropic system with a metric $ g_{\mu\nu} $ with positive spatial signature $ (-,+,+,+) $ filled by a perfect fluid with energy density~$ \rho $, isotropic pressure density~$ p $, 4-velocity~$ u^\mu $ and energy momentum tensor
	\begin{equation}
		T_{\mu\nu} = \rho u_\mu u_\nu + p h_{\mu\nu} ,
	\end{equation}
	where $ h_{\mu\nu} $ is the comoving spatial projector
	\begin{equation}
	h_{\mu\nu} = g_{\mu\nu} + u_\mu u_\nu ,
	\end{equation}
	and a uniform magnetic field with Faraday tensor~$ F_{\mu\nu} $.
	
	The time derivative of a generic tensor~$ T\indices{_\mu^\nu} $ is
	\begin{equation}
		\dot{T}\indices{_\mu^\nu} = u^\rho \nabla_\rho T\indices{_\mu^\nu} ,
	\end{equation}
	its spatial projected derivative
	\begin{equation}
		\D_\rho T\indices{_\mu^\nu} = h\indices{_\rho^\sigma} h\indices{_\mu^\alpha} h\indices{^\nu_\beta} \nabla_\sigma T\indices{_\alpha^\beta} ,
	\end{equation}
	the totally antisymmetric spatial tensor
	\begin{equation}
		\epsilon_{\mu\nu\rho} = \eta_{\mu\nu\rho\sigma} u^\sigma ,
	\end{equation}
	where $ \eta_{\mu\nu\rho\sigma} $ is the totally antisymmetric tensor with\linebreak $ {\eta^{0123} = 1/\sqrt{-g}} $,
	and the irreducible components of the velocity derivative are
	\begin{subequations}
		\begin{gather}
			\theta = \nabla_\mu u^\mu = \D_\mu u^\mu\\
			\sigma_{\mu\nu} = {\textstyle \frac{1}{2}} \left( \D_\mu u_\nu + \D_\nu u_\mu \right) - {\textstyle \frac{1}{3}} h_{\mu\nu} h^{\alpha\beta} \D_\alpha u_\beta\\
			\omega_{\mu\nu} = {\textstyle \frac{1}{2}} \left( \D_\mu u_\nu - \D_\nu u_\mu \right) , \quad \omega_\mu = {\textstyle \frac{1}{2}} \epsilon_{\mu\nu\rho} \omega^{\mu\rho}\\
			A_\mu = \dot{u}_\mu = u^\nu \nabla_\nu u_\mu .
		\end{gather}
	\end{subequations}
	
	It is now possible to describe the electromagnetic field in the Lorentz-Heaviside units: the electric field is  $ E_\mu = F_{\mu\nu} u^\nu $; the magnetic field is $ B^\mu = \epsilon^{\mu\nu\rho} F_{\nu\rho} / 2 $, with magnetic energy $ B^2 = B_\mu B^\mu $ and energy momentum tensor
	\begin{subequations}
		\begin{gather}
			T_{\mu\nu} = \frac{1}{2} B^2 u_\mu u_\nu + \frac{1}{6} B^2 h_{\mu\nu} + \Pi_{\mu\nu}\\
			\Pi_{\mu\nu} = \frac{1}{3} B^2 h_{\mu\nu} - B_\mu B_\nu .
		\end{gather}
	\end{subequations}
	The equations that describe our system are the Maxwell equations
	\begin{subequations}
		\begin{gather}
		\label{eq:basicEquations-BEvolution}
			\dot{B}_{\langle\mu\rangle} = \left( \sigma_{\mu\nu} + \epsilon_{\mu\nu\rho} \omega^\rho - \frac{2}{3} \theta h_{\mu\nu} \right) B^\nu\\
			\epsilon_{\mu\nu\rho} \D^\nu B^\rho = h\indices{_\mu^\nu} J_\nu - \epsilon_{\mu\nu\rho} A^\nu B^\rho\\
			\omega_\mu B^\mu = - \frac{1}{2} J_\mu u^\mu\\
		\label{eq:basicEquations-Bdiv}
			\D_\mu B^\mu = 0 ,
		\end{gather}
	\end{subequations}
	where $ J_\mu $ is the electric 4-current, and the projected Einstein equations
	\begin{subequations}
		\label{eq:basicEquations-Einstein}
		\begin{gather}
			R_{\mu\nu} u^\mu u^\nu = \frac{1}{2} (\rho + 3p + B^2)\\
			h\indices{_\mu^\nu} R_{\nu\rho} u^\rho = 0\\
			\begin{split}
				&h\indices{_\mu^\rho} h\indices{_\nu^\sigma} R_{\rho\sigma}\\
					&\qquad = \frac{1}{2} \left( \rho - p + \frac{1}{3} B^2 \right) h_{\mu\nu} + \Pi_{\mu\nu}
			\end{split}
		\end{gather}
	\end{subequations}
	in which $ R_{\mu\nu} $ is the Ricci tensor.
	
	The interaction between the fluid and the magnetic field is given by
	\begin{equation}
		\nabla^\mu T_{\mu\nu}^\mathrm{EM} = - F_{\mu\nu} J^\mu .
	\end{equation}	
	It is possible to use the Maxwell equation~\eqref{eq:basicEquations-BEvolution} to find the conservation law for the magnetic energy
	\begin{equation}
	\label{eq:basicEquations-magnetic-field-energy-conservation}
		\dot{(B^2)} = - \frac{4}{3} \theta B^2 - 2 \sigma_{\mu\nu} \Pi^{\mu\nu} ,
	\end{equation}
	and to derive the fluid energy conservation law from the temporal part of the Bianchi identities~$ u_\mu \nabla_\nu T^{\mu\nu} = 0 $
	\begin{equation}
	\label{eq:basicEquations-fluid-energy-conservation}
		\dot{\rho} = - (\rho + p) \theta ;
	\end{equation}
	from the spatial projected Bianchi identities~$ {h\indices{^\mu_\rho} \nabla_\nu T^{\rho\nu} = 0} $ it is possible to find the momentum conservation law
	\begin{equation}
		\begin{split}
			&\left( \rho + p + \frac{2}{3} B^2 \right) A_\mu\\
				&\qquad = - \D_\mu p - \epsilon_{\mu\nu\rho} B^\nu \epsilon^{\rho\alpha\beta} \D_\alpha B_\beta - \Pi_{\mu\nu} A^\nu
		\end{split}
	\end{equation}
	which, using
	\begin{equation}
		\epsilon_{\mu\nu\rho} B^\nu \epsilon^{\rho\alpha\beta} \D_\alpha B_\beta = \frac{1}{2} \D_\mu B^2 - B^\nu \D_\nu B_\mu ,
	\end{equation}
	gives
	\begin{equation}
	\label{eq:basicEquations-momentum-conservation}
		\begin{split}
			&\left( \rho + p + \frac{2}{3} B^2 \right) A_\mu = - \D_\mu p\\
				&\qquad - \frac{1}{2} \D_\mu B^2 + B^\nu \D_\nu B_\mu - \Pi_{\mu\nu} A^\nu .
		\end{split}
	\end{equation}

\section{Background model\label{sec:background}}
	We assume that our system is homogeneous and perturbed at first order by weak inhomogeneous perturbations. At the background level we have a homogeneous universe with an isotropic perfect fluid and a uniform magnetic field: such field cannot live with an isotropic metric, such as FRW, but it can be accommodated in an anisotropic model. We must use one of the Bianchi models because of the homogeneity and our model fits best in a Bianchi~I universe, which is the simplest anisotropic generalization of FRW, so our metric in synchronous gauge is
	\begin{equation}
		g_{\mu\nu} = \diag\left(-1, a_1^2 (t), a_2^2 (t), a_3^2 (t)\right) .
	\end{equation}
	
	These type of models were widely studied in literature in different assumptions and physical limits (see for example~\cite{bib:doroshkevich-1965,bib:thorne-1967,bib:jacobs-1969,bib:zeldovichNovikov-relativisticAstrophysics2TheStructureAndEvolutionOfTheUniverse,bib:king-2007}); we are here interested mainly in their behaviour after the matter-radiation equivalence, where the magnetic field can be reasonably small compared to the matter component. This regime was already studied in different works, for example by~\cite{bib:zeldovichNovikov-relativisticAstrophysics2TheStructureAndEvolutionOfTheUniverse} in radiation dominated universe; here we will recap~\cite{bib:barrow-1997}, which accounts for different type of anisotropic stresses in both radiation an matter dominated universe. We will, however, amend for their time behaviour in matter dominated universe and we will not neglect higher order corrections in the isotropic components.
	
	We assume that the magnetic field is oriented along the $ 3 $~axis, so the system is axisymmetric and $ a_1 = a_2 $; for\linebreak simplicity we call $ a = a_1 = a_2 $ and $ c = a_3 $. We have\linebreak $ u^\mu = (1,0,0,0) $.
	
	It is now straightforward to write the Einstein\linebreak equations~\eqref{eq:basicEquations-Einstein}
	\begin{subequations}
	\label{eqs:background-Einstein}
		\begin{gather}
		\label{eq:background-Einstein-00}
			2\ddotaovera{a} + \ddotaovera{c} = - \frac{1}{2} \left( \rho + 3p + B^2 \right)\\
		\label{eq:background-Einstein-11-22}
			\ddotaovera{a} + \hubble{a} \left( \hubble{a} + \hubble{c} \right) = \frac{1}{2} \left( \rho - p + B^2 \right)\\
		\label{eq:background-Einstein-33}
			\ddotaovera{c} + 2 \hubble{a} \hubble{c} = \frac{1}{2} \left( \rho - p - B^2 \right)
		\end{gather}
	\end{subequations}
	and the energy conservation laws for the system~\eqref{eq:basicEquations-fluid-energy-conservation} and~\eqref{eq:basicEquations-magnetic-field-energy-conservation}
	\begin{gather}
	\label{eq:background-fluid-energy-conservation}
		\dot{\rho} + \left(  2\hubble{a} + \hubble{c} \right) (\rho + p) = 0\\
	\label{eq:background-magnetic-field-energy-conservation}
		\dot{(B^2)} + 4 \hubble{a} B^2 = 0 .
	\end{gather}
	
	We define the Alfvén velocity, which is the energy ratio between magnetic field and fluid (note the factor $ 1/2 $ which differs from the usual definition)
	\begin{equation}
	\label{eq:background-definition-vA2}
		v_A^2 = \frac{B^2/2}{\rho} ,
	\end{equation}
	witch is responsible for the intensity of the anisotropies, the isotropic expansion~$ H $ and the anisotropy parameter~$ S $
	\begin{equation}
	\label{eq:background-definition-HS}
		3 H = 2 \hubble{a} + \hubble{c}, \quad S = \frac{1}{H} \left( \hubble{a} - \hubble{c} \right) .
	\end{equation}
	If we now assume a barotropic fluid with equation of state $ p = w \rho $ and $ w = \const $ the Einstein equation~\eqref{eq:background-Einstein-00} becomes
	\begin{equation}
	\label{eq:background-Einstein-00-HvA2S}
		3 \dot{H} + H^2 \left( 3 + \frac{2}{3} S^2 \right) = - \left[ \frac{1}{2} (1+3w) + v_A^2 \right] \rho ,
	\end{equation}
	subtracting equation~\eqref{eq:background-Einstein-33} from equation~\eqref{eq:background-Einstein-11-22} we get
	\begin{equation}
	\label{eq:background-S-evolution}
		H \dot{S} + \dot{H} S + 3 H^2 S = 2 v_a^2 \rho
	\end{equation}
	and summing 2~times equation~\eqref{eq:background-Einstein-11-22} to equation~\eqref{eq:background-Einstein-33} we eventually have
	\begin{equation}
	\label{eq:background-Einstein-trace-HvA2S}
		3 \dot{H} + 9 H^2 = \left[ \frac{3}{2} (1-w) + v_A^2 \right] \rho .
	\end{equation}
	From the definition of $ v_A^2 $ \eqref{eq:background-definition-vA2} and from the energy conservations \eqref{eq:background-fluid-energy-conservation} and \eqref{eq:background-magnetic-field-energy-conservation} we have
	\begin{equation}
	\label{eq:background-vA2-evolution}
		\dot{(v_A^2)} = \frac{\dot{(B^2)}/2 - \dot{\rho} v_A^2}{\rho} = v_A^2 H \left( 3w -1 -\frac{4}{3}S \right) .
	\end{equation}
	
	If we now assume that the magnetic field energy is small compared to the fluid energy we have $ v_A^2 \ll 1 $ and if we write
	\begin{equation}\label{eq:background-FRWSolution}
		H = H_{(0)} + H_{(1)} ,\quad \rho = \rho_{(0)} + \rho_{(1)}
	\end{equation}
	with $ H_{(1)} , \rho_{(1)} = \order{v_A^2} $ it is easy to see from equations \eqref{eq:background-Einstein-00-HvA2S} and \eqref{eq:background-Einstein-trace-HvA2S} that at $ 0 $-order in~$ v_A^2 $ we recover FRW and we have
	\begin{equation}
		H_{(0)} = \frac{2}{3(1+w) t} ,\quad \rho_{(0)} = 3 H_{(0)}^2 ,\quad S_{(0)} = 0 .
	\end{equation}
	
	The anisotropy is described by~$ S $ and equation~\eqref{eq:background-S-evolution} becomes at first order in~$ v_A^2 $
	\begin{equation}
	\label{eq:background-order-vA2-S}
		\dot{S} + \frac{1-w}{1+w} \frac{S}{t} = \frac{4}{1+w} \frac{v_A^2}{t} ,
	\end{equation}
	while equation~\eqref{eq:background-vA2-evolution} gives
	\begin{equation}
	\label{eq:background-order-vA2-vA2}
		\dot{(v_A^2)} = - \frac{2}{3} \frac{1-3w}{1+w} \frac{v_A^2}{t} .
	\end{equation}
	The isotropic part is contained in equations \eqref{eq:background-Einstein-00-HvA2S} and \eqref{eq:background-Einstein-trace-HvA2S}, which form a system whose solution is
	\begin{gather}
	\label{eq:background-order-vA2-rho}
		\rho_{(1)} = \frac{4}{1+w} \frac{H_{(1)}}{t} - \frac{4}{3 (1+w)^2} \frac{v_A^2}{t^2}\\
	\label{eq:background-order-vA2-H}
		\dot{H}_{(1)} + 2 \frac{H_{(1)}}{t} = - \frac{2}{9} \frac{1-3w}{(1+w)^2} \frac{v_A^2}{t^2} .
	\end{gather}
	
	We are interested only in anisotropies caused by the\linebreak magnetic field so we will put to~$ 0 $ the homogeneous solution of each equation, with the exception of~\eqref{eq:background-order-vA2-vA2}.
	
	\subsection{Radiation dominated universe\label{ssec:background-radiationDominatedUniverse}}
		For radiation dominated universe $ w = 1/3 $ and equation~\eqref{eq:background-order-vA2-vA2} gives
		\begin{equation}
			v_A^2 = v_{A0}^2 = \const .
		\end{equation}
		Equation~\eqref{eq:background-order-vA2-S} then gives
		\begin{equation}
			S = 6 v_A^2 = 6 v_{A0}^2 .
		\end{equation}
		
		From equation~\eqref{eq:background-order-vA2-rho} we get $ \rho $.
		
		From the definitions~\eqref{eq:background-definition-HS} we can get the values of $ a $ and $ c $. Finally we have
		\begin{gather}
			v_A^2 = v_{A0}^2 = \const , \quad t_0 = \const\\
			a \sim \left( \frac{t}{t_0} \right)^{1/2} \left( 1 + v_{A0}^2 \ln \left( \frac{t}{t_0} \right) \right)\\
			c \sim \left( \frac{t}{t_0} \right)^{1/2} \left( 1 - 2 v_{A0}^2 \ln \left( \frac{t}{t_0} \right) \right)\\
			\label{eq:background-solutionRadiationH} H = \frac{1}{2t}\\
			\rho = \frac{3}{4 t^2} (1 - v_{A0}^2) .
		\end{gather}
		
	\subsection{Matter dominated universe\label{ssec:background-matterDominatedUniverse}}
		For matter dominated universe $ w = 0 $ and eq.~\eqref{eq:background-order-vA2-vA2} gives
		\begin{equation}
			v_A^2 = v_{A0}^2 \left( \frac{t}{t_0} \right)^{-2/3} , \quad v_{A0}^2, t_0 = \const .
		\end{equation}
		From equation~\eqref{eq:background-order-vA2-S} we get
		\begin{equation}
			S(t) = 12 v_A^2 (t) .
		\end{equation}
		
		For the isotropic part we proceed as before: eq.~\eqref{eq:background-order-vA2-H} gives
		\begin{equation}
			H_{(1)} = - \frac{2}{3} \frac{v_A^2 (t)}{t}
		\end{equation}
		From equation~\eqref{eq:background-order-vA2-rho} we get $ \rho $.
		
		From the definitions~\eqref{eq:background-definition-HS} we can get the values of $ a $ and $ c $. Finally we have
		\begin{gather}
			v_A^2 = v_{A0}^2 \left( \frac{t}{t_0} \right)^{-2/3}\\
			a \sim \left( \frac{t}{t_0} \right)^{2/3} - 3 v_{A0}^2\\
			c \sim \left( \frac{t}{t_0} \right)^{2/3} + 9 v_{A0}^2\\
			H = \frac{2}{3t} (1 - v_A^2 (t))\\
		\label{eq:backgound-solutionMatter-rho}
			\rho = \frac{4}{t^2} \left( \frac{1}{3} - v_A^2 (t) \right) .
		\end{gather}

\section{Perturbed equations\label{sec:perturbedEquations}}
	We perturb all the quantities that govern our system while keeping synchronous gauge, thus the perturbed metric is
	\begin{subequations}
		\begin{gather}
			g_{\mu\nu} = g_{\mu\nu}^\back + \variation{g_{\mu\nu}}\\
			\variation{g_{\mu 0}} = 0 ,
		\end{gather}
	\end{subequations}
	where $ \back $ means the background value; we can define
	\begin{subequations}
		\begin{gather}
			\gamma_{\mu\nu} = \variation{g_{\mu\nu}}\\
			g_{\mu\rho} g^{\rho\nu} = \delta\indices{_\mu^\nu} \implies \variation{g^{\mu\nu}} = - \gamma^{\mu\nu} ,
		\end{gather}
	\end{subequations}
	where the indices of~$ \gamma_{\mu\nu} $ are raised and lowered with the unperturbed metric~$ g_{\mu\nu}^\back $. In the following we write the trace of $ \gamma_{\mu\nu} $ as $ {\gamma = \gamma\indices{_k^k}} $. The fluid velocity perturbation is
	$ \variation{u^\mu} $, with
	\begin{equation}
		u_\mu u^\mu = -1 \implies \variation{u^0} = 0 .
	\end{equation}
	The fluid energy perturbation is $ \variation{\rho} $ and the fluid pressure perturbation is $ \variation{p} = v_S^2 \variation{\rho} $; it holds
	\begin{subequations}
		\begin{gather}
			\label{eq:perturbedEquations-wEvolution} \dot{w} = - 3H (1+w) (v_S^2 - w)\\
			\label{eq:perturbedEquations-vs2const} w = \const \implies v_S^2 = w ,
		\end{gather}
	\end{subequations}
	but we keep $ {v_S^2} $ as an arbitrary function and possibly different from~$ w $; the reason of this choice will be clear in section~\ref{ssec:solutionProcedure-matterSmallScales}.
	
	The perturbed magnetic field must remain pure spatial at all orders, as shown in appendix~\ref{app:magneticField}, so the condition $ B_\mu u^\mu = 0 $ holds at all perturbative orders and the perturbation to the magnetic field satisfies
	\begin{subequations}
		\begin{gather}
			\label{eq:perturbedEquations-magneticPerturbations} \variation{(B_\mu B^\mu)} = \variation{(B^2)} = \gamma_{33} B^3 B^3 + 2 c^2 \variation{B^3} B^3 \\
			B_\mu u^\mu = 0 \implies \variation{B^0} = c^2 B^3 \variation{u^3} .
		\end{gather}
	\end{subequations}
	
	Accordingly to~\cite{bib:landau-classicalFields,bib:weinberg-gravitationAndCosmology} the perturbed Christoffel symbols are
	\begin{equation}
		\variation{\Gamma_{\mu\nu}^\rho} = \frac{1}{2} g_\back^{\rho\sigma} \left( \nabla_\mu^\back \gamma_{\nu\sigma} + \nabla_\nu^\back \gamma_{\mu\sigma} - \nabla_\sigma^\back \gamma_{\mu\nu} \right)
	\end{equation}
	and the perturbed Ricci tensor is
	\begin{equation}
		\variation{R_{\mu\nu}} = \nabla_\rho^\back \variation{\Gamma_{\mu\nu}^\rho} - \nabla_\nu^\back \variation{\Gamma_{\mu\rho}^\rho} .
	\end{equation}
	
	We are now ready to perturb the exact equations of section~\ref{sec:basicEquations}. We notice that, because of the homogeneity of the background model, when applied to the perturbation of a scalar quantity the comoving time derivative~$ \protect\dot{s} $ is the same as the synchronous time derivative~$ \partial_0 s $, so we make no difference between them in the following. The fluid energy conservation~\eqref{eq:basicEquations-fluid-energy-conservation} becomes
	\begin{equation}
	\label{eq:perturbedEquations-fluidEnergyConservation}
		\begin{split}
			&\dot{\variation{\rho}} + \left(2\hubble{a} + \hubble{c}\right) \left( \variation{\rho} + \variation{p} \right)\\
				&\qquad + (\rho^\back + p^\back) \left( \partial_i \variation{u^i} + \frac{1}{2} \dot{\gamma} \right) = 0
		\end{split}
	\end{equation}
	and the magnetic field energy conservation~\eqref{eq:basicEquations-magnetic-field-energy-conservation} gives
	\begin{equation}
	\label{eq:perturbedEquations-magneticFieldEnergyConservation}
		\begin{split}
			&\dot{(\variation{(B^2)})} + 4\hubble{a} \variation{(B^2)} + 2 B^{2(0)}\\
				&\qquad \cdot \left( \partial_i \variation{u^i} - \partial_3 \variation{u^3} + \frac{1}{2} \dot{\gamma} - \frac{1}{2} \dot{\gamma}\indices{_3^3} \right) = 0 .
		\end{split}
	\end{equation}
	
	The Einstein $ 00 $~equation is (we will always use Einstein equations with a lower and an upper index)
	\begin{equation}
	\label{eq:perturbedEquations-Einstein00}
		\begin{split}
			&\frac{1}{2} \ddot{\gamma} + \hubble{a} \dot{\gamma} - \left( \hubble{a} - \hubble{c} \right) \dot{\gamma}\indices{_3^3}\\
				&\qquad + \frac{1}{2} \left( \variation{\rho} + 3 \variation{p} \right) + \frac{1}{2} \variation{(B^2)} = 0 ,
		\end{split}
	\end{equation}
	while the $ 33 $~equation reads
	\begin{equation}
	\label{eq:perturbedEquations-Einstein33}
		\begin{split}
			&\partial_k \partial^3 \gamma\indices{_3^k} - \frac{1}{2} \left( \partial_k \partial^k \gamma\indices{_3^3} + \partial_3 \partial^3 \gamma \right)\\
				&\qquad + \frac{1}{2} \ddot{\gamma}\indices{_3^3} + \frac{1}{2} \left(2\hubble{a} + \hubble{c}\right) \dot{\gamma}\indices{_3^3} + \frac{1}{2} \hubble{c} \dot{\gamma}\\
				&\qquad - \frac{1}{2} \left( \variation{\rho} - \variation{p} \right) + \frac{1}{2} \variation{(B^2)} = 0 ;
		\end{split}
	\end{equation}
	to remove $ \partial_3 \partial^k \gamma\indices{_k^3} $ from the last equation we need to use the derivative of the $ 03 $~equation with respect to the $ 3 $~index
	\begin{equation}
	\label{eq:perturbedEquations-Einstein03}
		\begin{split}
			&\partial_0 \left( \partial_3 \partial^k \gamma\indices{_k^3} \right) - \partial_3 \partial^3 \dot{\gamma} + 2 \hubble{a} \partial_3 \partial^k \gamma\indices{_k^3}\\
				&\qquad - \left( \hubble{a} - \hubble{c} \right) \partial_3 \partial^3 \gamma - \left( \hubble{a} - \hubble{c} \right) \partial_3 \partial^3 \gamma\indices{_3^3}\\
				&\qquad = - 2 (\rho^\back + p^\back) \partial_3 \variation{u^3} .
		\end{split}
	\end{equation}
	If we had used equations~\eqref{eq:basicEquations-Einstein} we would have found the same results.
	
	By imposing the null divergence of the magnetic field~\eqref{eq:basicEquations-Bdiv} we get
	\begin{equation}
	\label{eq:perturbedEquations-magneticFieldDivergence}
		\partial_i \variation{B^i} + \frac{1}{2} B^3 \partial_3 \gamma = 0 .
	\end{equation}
	
	The last equation we need is the conservation of the momentum~\eqref{eq:basicEquations-momentum-conservation} (note that $ A^\mu $ has only the first order component): we define an index $ P \in \{1, 2\} $ that lies on the plane orthogonal to the background magnetic field and we write the divergence of the momentum conservation on the $ 12 $-plane ($ \partial_1 ()^1 + \partial_2 ()^2 $)
	\begin{equation}
	\label{eq:perturbedEquations-momentumConservationDerivativeP}
		\begin{split}
			&(\rho^\back + p^\back) \left( \partial_0 \partial_P \variation{u^P} + 2 \hubble{a} \partial_P \variation{u^P} \right)\\
				&\qquad + B_\back^{2} \left[ \partial_0 \partial_P \variation{u^P} + \left(2 \hubble{a} + \hubble{c} \right) \partial_P \variation{u^P} \right]\\
				&\qquad + \partial_P \variation{u^P} \partial_0 \left( p^\back + \frac{1}{2} B_\back^{2} \right)\\
				&\qquad + \partial_P \partial^P \left( \variation{p} + \frac{1}{2} \variation{(B^2)} \right) - B^3 \partial_3 \partial_P \variation{B^P}\\
				&\qquad + B_\back^{2} \left( \frac{1}{2} \partial_P \partial^P \gamma\indices{_3^3} - \partial_P \partial^3 \gamma\indices{_3^P} \right) = 0
		\end{split}
	\end{equation}
	and the derivative of the $ 3 $~component along the $ 3 $~axis
	\begin{equation}
	\label{eq:perturbedEquations-momentumConservationDerivative3}
		\begin{split}
			&(\rho^\back + p^\back) \left( \partial_0 \partial_3 \variation{u^3} + 2\hubble{c} \partial_3 \variation{u^3} \right)\\
				&\qquad + \partial_3 \variation{u^3} \partial_0 \left( p^\back + \frac{1}{2} B_\back^{2} \right)\\
				&\qquad + \partial_3 \partial^3 \left( \variation{p} + \frac{1}{2} \variation{(B^2)} \right) + 2 \hubble{a} B_\back^{2} \partial_3 \variation{u^3}\\
				&\qquad - B^3 \partial_3 \partial_3 \variation{B^3} - \frac{1}{2} B_\back^{2} \partial_3 \partial^3 \gamma\indices{_3^3} = 0 .
		\end{split}
	\end{equation}
	
	The system~\eqref{eq:perturbedEquations-fluidEnergyConservation}-\eqref{eq:perturbedEquations-momentumConservationDerivative3} fully characterizes the evolution of the perturbed quantities and it is the ground of the following analysis. Compared to~\cite{bib:tsagas-2000} we fully related the background anisotropy to the magnetic field, without the need of additional hypothesis.

\section{Gauge Modes\label{sec:gaugeModes}}
	Fixing the synchronous gauge does not end the freedom of coordinate choice: we can still make a gauge transformation preserving the synchronous gauge.
	
	We follow the same scheme as of~\cite{bib:montani-primordialCosmology}: we make a generic coordinate transformation of the form
	\begin{equation}
		x^\mu \rightarrow x^\mu + \epsilon^\mu
	\end{equation}
	with small~$ \epsilon^\mu $ and we keep terms up to~$ \order{\epsilon} $.
	
	The metric tensor becomes
	\begin{equation}
		g'_{\mu\nu} (x') = g_{\mu\nu} (x) - g_{\mu\sigma} (x) \partial_\nu \epsilon^\sigma - g_{\rho\nu} (x) \partial_\mu \epsilon^\rho .
	\end{equation}
	If we define
	\begin{equation}
	\label{eq:gaugeModes-trasfGauge-gmn}
		\begin{split}
			&\Delta g_{\mu\nu} = g'_{\mu\nu} (x) - g_{\mu\nu} (x) = - g_{\mu\lambda} (x) \partial_\nu \epsilon^\lambda\\
			&\qquad\qquad - g_{\lambda\nu} (x) \partial_\mu \epsilon^\lambda - \epsilon^\lambda \partial_\lambda g_{\mu\nu} (x)\\
			&\qquad = - \nabla_\mu \epsilon_\nu - \nabla_\nu \epsilon_\mu
		\end{split}
	\end{equation}
	to preserve the synchronous gauge we need
	$ \Delta g_{0\mu} = 0 $ which gives $ \epsilon^0 = \epsilon^0 (x^j) $ and
	\begin{subequations}
		\begin{gather}
			\epsilon^P = \tilde{\epsilon}^P (x^j) + \partial^P \epsilon^0 (x^j) \, a^2 \int \frac{\dd{t}}{a^2} ,\\
			\epsilon^3 = \tilde{\epsilon}^3 (x^j) + \partial^3 \epsilon^0 (x^j) \, c^2 \int \frac{\dd{t}}{c^2} ,
		\end{gather}
	\end{subequations}
	where $ \epsilon_0 (x^j) $ and $ \tilde{\epsilon}^i (x^j) $ are arbitrary functions of the spatial coordinates: we still have 4 unused degrees of freedom represented by the functions~$ \epsilon^0 $ and~$ \tilde{\epsilon}^i $.
	
	If we take the functions~$ \epsilon^0 $ and~$ \tilde{\epsilon}^i $ of the same order of the perturbations then the transformation given by equation~\eqref{eq:gaugeModes-trasfGauge-gmn} can be seen both as a gauge transformation and as a transformation of the functions~$ \gamma_{\mu\nu} $ within fixed synchronous gauge: in the latter case equation~\eqref{eq:gaugeModes-trasfGauge-gmn} gives the value of~$ \Delta \gamma_{\mu\nu} $. In the same way the stress-energy tensor transforms as
	\begin{equation}
		\begin{split}
			&\Delta T_{\mu\nu} = - T_{\mu\lambda} \partial_\nu \epsilon^\lambda - T_{\lambda\nu} \partial_\mu \epsilon^\lambda - \epsilon^\lambda \partial_\lambda T_{\mu\nu}\\
				&\qquad = - T_{\mu\lambda} \nabla_\nu \epsilon^\lambda - T_{\lambda\nu} \nabla_\mu \epsilon^\lambda - \epsilon^\lambda \nabla_\lambda T_{\mu\nu}
		\end{split}
	\end{equation}
	and if we see these as transformations on the physical variables instead of the coordinates we obtain the gauge modes for~$ \variation{T_{\mu\nu}} $. Substituting the explicit expression of~$ T_{\mu\nu} $ as the sum of the fluid and the magnetic field components we see that the transformation acts separately on the two components and we get for the fluid density perturbation
	\begin{equation}
	\label{eq:gaugeModes-diff-dr}
		\begin{split}
			&\Delta \variation{\rho} = - \epsilon^0 \dot{\rho}^\back\\
				&\qquad = 3H (\rho^\back + p^\back) \epsilon^0 = 3H (1+w) \rho^\back \epsilon^0 .
		\end{split}
	\end{equation}
	
	In section~\ref{sec:perturbedEquations} we linearised the equations and so the gauge transformations solve our equations and we call them gauge perturbations or gauge modes: these solutions are not physical because they correspond to a simple change in the reference frame. We are looking for a physical solution for the time dependence of~$ \variation{\rho} $ so the most interesting gauge transformation is given by equation~\eqref{eq:gaugeModes-diff-dr}.
	
	Having the knowledge of gauge modes it is possible to construct gauge invariant variables, in a similar way as done in~\cite{bib:noh-1995}. We have
	\begin{equation}
		\Delta \variation{u^i} = \partial^i \epsilon^0
	\end{equation}
	so our main scalar variable should be
	\begin{subequations}
	\begin{gather}
		\variation{\rho}^\mathrm{GI} = \partial^i \partial_i \variation{\rho} - 3H (1+w) \rho^\back \partial_i \variation{u^i}\\
		\Delta \variation{\rho}^\mathrm{GI} = 0 .
	\end{gather}
	\end{subequations}
	It is easy to check that it is exactly the variable used in~\cite{bib:barrow-2007}, expressed in synchronous gauge. We will, however, not need it because the vorticity part~$ H \partial_i \variation{u^i} $ decays in time with respect to~$ \partial_i \partial^i\variation{\rho} / \rho^\back $ and we are interested in late time dynamics. We will also not need the laplacian, because we'll use Fourier expansions so it will reduce to a multiplicative term: for late times we can assume~$ \variation{\rho} $ to be gauge invariant.
	
	It is possible to watch this approximation from another perspective, shown in Appendix~\ref{app:lateTimesGauge}.

\section{Analytical Solutions\label{sec:solutionProcedure}}
	If we write the perturbations as Fourier transforms we see that the system imposes different evolution to the perturbations that propagates along the background magnetic field, with $ {\partial_P (\dots) = 0} $, and the perturbations that propagates orthogonally to the background magnetic field, with\linebreak $ {\partial_3 (\dots) = 0} $. These different modes are however coupled by the magnetic stress-energy tensor tensorial nature.
	
	To simplify the equations we use the barotropic state equation for the fluid, so $ {p^\back = w \rho^\back} $ with $ {w = \const} $ and\linebreak $ {\variation{p} = v_S^2 \variation{\rho}} $, and the Fourier expansion for the spatial part of the perturbations, so the spatial dependence is of the form $ {\nepernumber^{\iunit k_j x^j}} $. We define the new variables
	\begin{gather}
		\Delta = \frac{\variation{\rho}}{(1+w) \rho^\back}\\
		G = \frac{1}{2} \gamma\\
		T = \frac{1}{2} \gamma\indices{_3^3}\\
		M = \frac{\variation{(B^2)}}{B_\back^{2}} .
	\end{gather}
	
	Our differential equation system is not simple but we can solve it for small magnetic fields by keeping only terms up to first order in~$ {v_A^2} $: we shall remember that $ {S} $ is already at first order while $ {\Delta} $, $ {G} $, $ {T} $ and $ {\variation{u^i}} $ have also a 0-order (FRW) part; $ {M} $ has only the 0-order part because it is always multiplied by $ {v_A^2} $ because $ {\variation{(B^2)} = B_\back^{2} M = 2 \rho^\back v_A^2 M} $. In the same way, looking at our system also $ {T} $ is always multiplied by~$ {v_A^2} $: this is because it does not affect density perturbations unless some anisotropy is present.
	
	We also use eq.~\eqref{eq:perturbedEquations-vs2const} to discard terms proportional\linebreak to~$ {w - v_S^2} $ or to~$ \dot{(v_S^2)} $, unless multiplied by $ {k_i k^i} $ or $ {k_3 k^3} $. This is because, while they are equal to~$ 0 $ for~$ w=\const $, we will need them in sec.~\ref{ssec:solutionProcedure-matterSmallScales}.
	
	The fluid energy conservation equation~\eqref{eq:perturbedEquations-fluidEnergyConservation} in the new variables reads
	\begin{equation}
	\label{eq:solutionProcedure-G}
		\dot{G} = - \dot{\Delta} - \partial_i \variation{u^i} .
	\end{equation}
	
	Similarly we rewrite the magnetic energy conservation \eqref{eq:perturbedEquations-magneticFieldEnergyConservation}
	\begin{equation}
	\label{eq:solutionProcedure-M}
		\begin{split}
			&\dot{M} = -2 \left( \partial_P \variation{u^P} + \dot{G} - \dot{T} \right)\\
				&\qquad = 2 \left( \dot{\Delta} + \dot{T} + \partial_3 \variation{u^3} \right) ,
		\end{split}
	\end{equation}
	where we found the last equality by using the fluid energy conservation.
	
	Combining Einstein $ 33 $~equation~\eqref{eq:perturbedEquations-Einstein33} with its derivative with respect to time and using the derivative of Einstein $ 03 $~equation~\eqref{eq:perturbedEquations-Einstein03} with respect to the $ 3 $-index in order to take care of $ {\partial_i \partial^3 \gamma\indices{_3^i}} $ terms we get an equation for~$ {T} $. Because $ {T} $ only appears in the system in terms that are multiplied by~$ {v_A^2} $, we will only need this equation at 0-order:
	\begin{equation}
	\label{eq:solutionProcedure-Tord0}
		\begin{split}
			&3 (1+w) \dddot{T} + 10 \frac{\ddot{T}}{t} + 2 \frac{1-3w}{1+w} \frac{\dot{T}}{t^2} - 8 \frac{\partial_3 \variation{u^3}}{t^2}\\
				&\qquad + 2 \frac{\ddot{G}}{t} + \frac{2}{3} \frac{1-3w}{1+w} \frac{\dot{G}}{t^2}\\
				&\qquad - 2 \left(1-v_S^2\right) \frac{\dot{\Delta}}{t^2} + \frac{4}{3} (1-v_S^2) \frac{1+3w}{1+w} \frac{\Delta}{t^3}\\
				&\qquad + 3 (1+w) (k_i k^i \dot{T} - k_3 k^3 \dot{G}) = 0
		\end{split}
	\end{equation}
	
	We can use the fluid energy conservation equation~\eqref{eq:solutionProcedure-G} to eliminate~$ {G} $ from the other equations. This way the Einstein $ 00 $-equation~\eqref{eq:perturbedEquations-Einstein00} reads
	\begin{equation}
	\label{eq:solutionProcedure-Delta}
		\begin{split}
			&\ddot{\Delta} + 2 H \left( 1 + \frac{1}{3} S \right) \dot{\Delta} - \frac{1}{2} (1+3v_S^2) (1+w) \rho \Delta\\
				&\qquad + \partial_0 \partial_i \variation{u^i} + 2 H \left( 1 + \frac{1}{3} S \right) \partial_i \variation{u^i}\\
				&\qquad + \frac{4}{3 (1+w)} S \frac{\dot{T}}{t} - \frac{4}{3 (1+w)^2} v_A^2 \frac{M}{t^2} = 0 .
		\end{split}
	\end{equation}
	
	We obtain the evolution equation for the divergence of the 4-velocity by summing eqs.~\eqref{eq:perturbedEquations-momentumConservationDerivativeP} an~\eqref{eq:perturbedEquations-momentumConservationDerivative3}; we then use equation~\eqref{eq:perturbedEquations-Einstein33} to remove the $ {\partial_i \partial^3 \gamma\indices{_3^i}} $ term and equation~\eqref{eq:perturbedEquations-magneticFieldDivergence} to remove the divergence of the magnetic field. Doing so we find
	\begin{equation}
	\label{eq:solutionProcedure-ui}
		\begin{split}
			&\left(1 + \frac{2}{1+w} v_A^2 \right) \partial_0 \partial_i \variation{u^i} +\\
				&\quad + \left[ (2-3w) H +\left( \frac{v_A^2}{1+w} + \frac{1}{3} S \right) \frac{4}{3(1+w)} \frac{1}{t} \right] \partial_i \variation{u^i} =\\
					&\qquad = - v_S^2 \partial_i \partial^i \Delta- \frac{v_A^2}{1+w} \partial_i \partial^i M\\
					&\qquad + \frac{2}{1+w} v_A^2 \partial_0 \partial_3 \variation{u^3} + 2 \left( \frac{v_A^2}{1+w} + S \right) \frac{2}{3 (1+w)} \frac{\partial_3 \variation{u^3}}{t}\\
					&\qquad - \frac{2}{1+w} v_A^2 \left[ \ddot{T} + \frac{2}{1+w} \frac{\dot{T}}{t} + \frac{2}{3 (1+w)} \frac{\dot{G}}{t} \right]\\
					&\qquad + \frac{4}{3 (1+w)} (1-v_S^2) v_A^2 \frac{\Delta}{t^2} .
		\end{split}
	\end{equation}
	
	We will need also equation~\eqref{eq:perturbedEquations-momentumConservationDerivative3} which reads, using equation~\eqref{eq:perturbedEquations-magneticPerturbations} to remove $ {\partial_3 \variation{B^3}} $,
	\begin{equation}
	\label{eq:solutionProcedure-u3}
			\partial_0 \partial_3 \variation{u^3} + \left( 2-3w-\frac{4}{3}S \right) H \partial_3 \variation{u^3} + \partial_3 \partial^3 (v_S^2 \Delta) = 0 .
	\end{equation}
	
	Thus we restated the dynamical system~\eqref{eq:perturbedEquations-fluidEnergyConservation}-\eqref{eq:perturbedEquations-momentumConservationDerivative3} in a more suitable form which is more appropriate for the following analysis.
	
	\subsection{Radiation dominated universe at large scales\label{ssec:solutionProcedure-radiationLargeScales}}
		In radiation dominated universe we have~$ {w = v_S^2 = 1/3} $ and at large scales we can set $ {k^2 \approx k_3 k^3 \approx 0} $.
		It is easy to check that, once we get rid of the scale dependent terms, eq.~\eqref{eq:solutionProcedure-Tord0}, \eqref{eq:solutionProcedure-Delta} and~\eqref{eq:solutionProcedure-ui} reduces respectively to
		\begin{gather}
			2 \dddot{T} + 5 \frac{\ddot{T}}{t} - 4 \frac{\partial_3 \variation{u^3}}{t^2} - \frac{\partial_0 \partial_i \variation{u^i}}{t} - \frac{\ddot{\Delta}}{t} - \frac{2}{3} \frac{\dot{\Delta}}{t^2} + \frac{2}{3} \frac{\Delta}{t^3} = 0\\
			\begin{split}
				&\ddot{\Delta} + \left( 1 + 2 v_A^2 \right) \frac{\dot{\Delta}}{t} - (1 - v_{A0}^2) \frac{\Delta}{t^2} + 6 v_A^2 \frac{\dot{T}}{t} - \frac{3}{4} v_A^2 \frac{M}{t^2}\\
					&\qquad + \partial_0 \partial_i \variation{u^i} + \left( 1 + 2 v_A^2 \right) \frac{\partial_i \variation{u^i}}{t} = 0
			\end{split}\\
			\begin{split}
				&\left(1 + \frac{3}{2} v_A^2 \right) \partial_0 \partial_i \variation{u^i} + \frac{1 + 4 v_A^2}{2} \frac{\partial_i \variation{u^i}}{t} =\\
					&\qquad = \frac{3}{2} v_A^2 \partial_0 \partial_3 \variation{u^3} + \frac{27}{4} v_A^2 \frac{\partial_3 \variation{u^3}}{t}\\
					&\qquad  - \frac{3}{2} v_A^2 \ddot{T} - \frac{9}{4} v_A^2 \frac{\dot{T}}{t} + \frac{3}{4} v_A^2 \frac{\dot{\Delta}}{t} + \frac{1}{2} v_A^2 \frac{\Delta}{t^2} .
			\end{split}
		\end{gather}
		This system, together with~\eqref{eq:solutionProcedure-M} and \eqref{eq:solutionProcedure-u3}, is satisfied by a power law solution and could be reduced to a pure algebraic problem, but we found simpler to solve it for~$ {v_A^2 = 0} $ and then look perturbatively for the corrections in~$ {v_A^2} $. We found
		\begin{equation}
		\label{eq:solutionProcedure-largeScalesSolutionsComplete}
			\Delta = \frac{\Delta_\textrm{gauge}}{t} + \Delta_\textrm{grow} t^{1-2 v_{A0}^2} + \Delta_1 t^{1/2-2 v_{A0}^2}  + \Delta_2 t^{1/2+4 v_{A0}^2} .
		\end{equation}
		It can be shown that the $ t^{1/2} $~modes are related to a non-vanishing divergence of the background velocity\linebreak $ {\partial_i \variation{u^i} = i k_i \variation{u^i}} $: strictly speaking, we should have imposed the $ {k_i \approx 0} $ condition, thus finding only the $ t $ and $ 1/t $ modes:
		\begin{equation}
		\label{eq:solutionProcedure-largeScalesSolutions}
			\Delta = \frac{\Delta_\textrm{gauge}}{t} + \Delta_\textrm{grow} t^{1-2 v_{A0}^2}
		\end{equation}
		and recovering the usual FRW solution in the limit~$ {v_A^2 \rightarrow 0} $.
		
		Using~\eqref{eq:gaugeModes-diff-dr} and~\eqref{eq:background-solutionRadiationH} we find that $ {1/t} $ is a gauge mode, while $ {t^{1-2 v_{A0}^2}} $ is the physical growing mode, with the correction due to the magnetic field.
		
		We find our solution simpler than the one of~\cite{bib:barrow-2007}, and with a clearer physical interpretation of the solutions, but our physical growing mode follows a slightly different temporal law, although this correction is small given~$ {v_A^2} \ll 1 $. We also find simpler the comparison of our solution with the non magnetic one of~\cite{bib:weinberg-gravitationAndCosmology}.
		
		We see in~\eqref{eq:solutionProcedure-largeScalesSolutions} that the magnetic field reduces the growing rate of density perturbations, but by an amount of order~$ {\order{v_A^2} \ll 1} $. This effect has long been known, and it is due to the extra magnetic pressure. A similar behavoiur was found in~\cite{bib:barrow-2007} and~\cite{bib:tsagas-2000}, although with the differences stated above.
		
		We finally note a difference between our solution~\eqref{eq:solutionProcedure-largeScalesSolutionsComplete} and the one of~\cite{bib:barrow-2007}: the non dominant mode is~$ t^{1/2} $ in our formalism, while~$ t^{-1/2} $ in their. At a more careful analysis, our equations tend correctly to the ones of~\cite{bib:weinberg-gravitationAndCosmology} for~$ {v_A^2 \rightarrow 0} $ and we obtain in such limit the same solutions of~\cite{bib:kolb-turner,bib:montani-primordialCosmology}, including the $ t^{1/2} $~mode. Such discrepancy is therefore between the synchronous and covariant formalisms, and it is besides the purposes of our paper.
		
	\subsection{Matter dominated universe at small scales\label{ssec:solutionProcedure-matterSmallScales}}
		In this section we analyse the perturbations in a matter dominated universe ($ {w=0} $), in the regime in which the\linebreak anisotropies are small with respect to the background. We expand in Fourier the spatial part of each quantity like $ \mathrm{e}^{\mathrm{i} k_j x^j} $, with $ {k_j = \const} $, and we define $ {k^2 = k_i k^i} $.
			
		Being at small scales means $ {k^2 \gg H^2} $ and assuming\linebreak $ {v_S^2, v_A^2 \ll 1} $ we can greatly simplify our equations, keeping only terms in $ {v_S^2} $ or $ {v_A^2} $ that are multiplied by $ {k^2} $ and dropping terms of order $ {v_S^2} $ and $ {v_A^2} $. This means that the effect of the sound speed and the Alfvén speed is relevant only at very small scales, as we will see from the solutions of our equations. This approximation, although still relativistic and so comparable to other result in literature, for example~\cite{bib:barrow-2007}, will give the nonrelativistic limit, as shown in section~\ref{sssec:solutionProcedure-matterSmallScales-analyticalSolutions}
		
		\subsubsection{Sound speed and Alfvén speed}
			First we need some considerations regarding the sound\linebreak speed. From a formal point of view, the sound speed is related to the barotropic index~$ {w} $ by~\eqref{eq:perturbedEquations-wEvolution} and $ {w = \const} $ implies $ {v_S^2 = w} $, so it should vanish. From a physical point of view we need a nonvanishing sound speed and we can also estimate its value. While formally the best solution to this problem would be using a two fluid model, with a different equation of state for perturbations, here we will simply drop the relation between $ {v_S^2} $ and $ {w} $ and assume that the perturbed fluid follows a different equation of state with respect to the background fluid. This is correct in the Newtonian approximation and it's in fact the standard way of handling things~\cite{bib:weinberg-gravitationAndCosmology,bib:lattanzi-2012}, while putting $ {v_S^2 = 0} $ at the end will recover the full covariant value of our calculations for studying pure magnetic effects.
			
			We proceed as in~\cite{bib:weinberg-gravitationAndCosmology}: we use an adiabatic sound speed
			\begin{equation}
				v_S^2 = \frac{\variation{p}}{\variation{\rho}} \sim \frac{\gamma p}{\rho} \sim \rho^{\gamma-1} \sim t^{2(1-\gamma)}
			\end{equation}
			where $ {\gamma} $~is the heat ratio. We write $ {\nu = \gamma - 4/3 \geq 0} $ so
			\begin{equation}
				v_S^2 = v_{S0}^2 \left(\frac{t}{t_0}\right)^{-2\left(\nu + \frac{1}{3}\right)} .
			\end{equation}
			
			We can estimate more precisely the sound speed value, and it's possible to show that the adiabatic sound speed is \cite{bib:weinberg-gravitationAndCosmology,bib:lattanzi-2012}				
			\begin{equation}
				\left. v_S^2 \right|_{z<z_\mathrm{rec}} = \frac{1}{3} \frac{k_B T_b \sigma}{m_p + k_B T_b \sigma} ,\ \left. v_S^2 \right|_{z>z_\mathrm{rec}} = \frac{5}{3} \frac{k_B T_b}{m_p},
			\end{equation}
			where $ {z_\mathrm{rec}} $~is the redshift value at recombination, $ {k_b} $~is the Boltzmann constant, $ {T_b} $~is the baryons temperature,\linebreak $ {m_p} $~is the proton mass and $ {\sigma} $~is the specific entropy, whose value is $ {\sigma = 4 a_\mathrm{SB} T^3 / 3 n_b k_B \approx 1.5 \cdot 10^9} $, being $ {a_\mathrm{SB}} $~the Stefan–Boltzmann constant and $ {T} $~the gas temperature. We neglected any anisotropic effects in temperature, because they would be related to the next order corrections. The baryons temperature is the same of the photons until~$ z \approx 100 $, due to residual Thomson scattering, and decreases faster thereafter:
			\begin{subequations}
				\begin{gather}
					\left. T_b \right|_{z>100} = T_\gamma = \left. T_\gamma \right|_{z=0} (1+z) ,\ \left. T_\gamma \right|_{z=0} \approx \SI{2.7}{\kelvin}\\
					\left. T_b \right|_{z<100} \propto (1+z)^2 .
				\end{gather}
			\end{subequations}
			
\begin{figure}[H]
				\centering
				\includegraphics[width=\columnwidth,keepaspectratio]{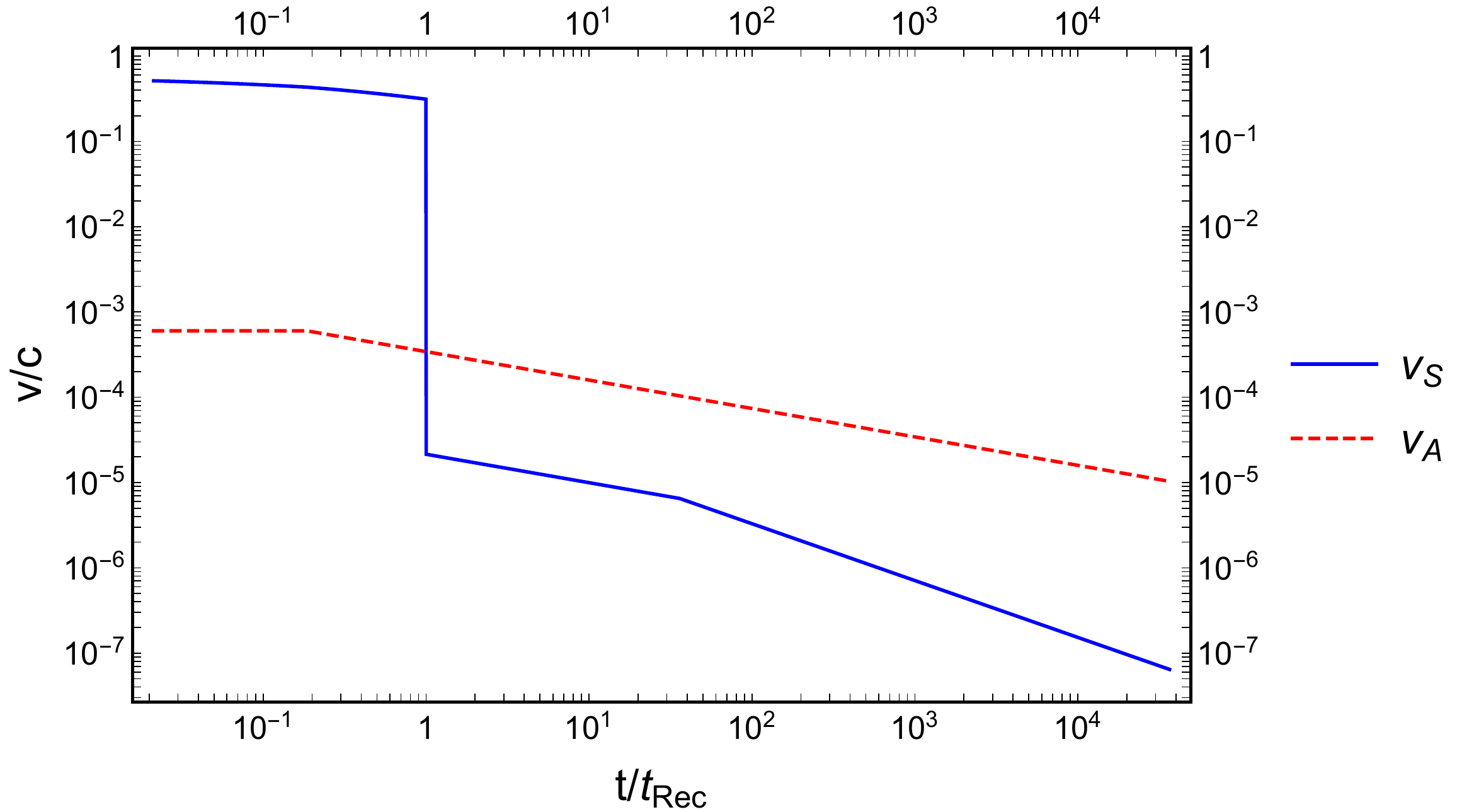}
				\caption{Plot of the sound speed and the Alfvén speed. We see that sound speed dominates until recombination, where suddenly the Alfvén velocity becomes important.\label{fig:velocities}}
			\end{figure}

			Comparing the two expressions we see that right after\linebreak recombination and until complete decoupling, so for\linebreak $ {z_\mathrm{rec} = 1100 > z > 100 = z_\mathrm{dec}} $, we have~$ {\nu = 0} $ and the cosmic medium behaves like a nonrelativistic fluid with $ {\gamma = 4/3} $: the total energy density is dominated by hydrogen rest mass but the pressure is dominated by radiation. After the end of Thomson scattering effects and until reionization, for $ 100 > z > 10 $, $ \nu \simeq 1/3 $ and the cosmic medium behave like a relativistic fluid with~$ {\gamma \simeq 5/3} $. The plot of the sound speed and of the Alfvén speed is in fig.~\ref{fig:velocities}.
			
			We define two constants addressing the effect of sound speed and Alfvén speed after recombination. Taking the time dependence of $ {k^2} $ depending only on the $ 0 $-order part of the background metric because it always appears multiplied by $ {v_S^2} $ or $ {v_A^2} $, we have respectively
			\begin{equation}\label{eq:solutionProcedure-LambdaConstants}
				\Lambda_S^2 = v_S^2 k^2 t^{2\gamma - 2/3} ,\quad \Lambda_A^2 = v_A^2 k^2 t^2 .
			\end{equation}
			
			For a more detailed discussion about the sound speed see~\cite{bib:bonometto-1974}.
			
		\subsubsection{Analytical solutions\label{sssec:solutionProcedure-matterSmallScales-analyticalSolutions}}
			Using the assumptions of section~\ref{ssec:solutionProcedure-matterSmallScales} we can greatly simplify our equations. The energy conservation~\eqref{eq:solutionProcedure-G} and the magnetic field energy conservation~\eqref{eq:solutionProcedure-M} retain the same form. The Einstein $ 00 $-equation~\eqref{eq:solutionProcedure-Delta} now reads
			\begin{equation}
				\ddot{\Delta} + \frac{4}{3 t} \dot{\Delta} - \frac{2}{3 t^2} \Delta + \partial_0 \partial_i \variation{u^i} + \frac{4}{3 t} \partial_i \variation{u^i} = 0 .
			\end{equation}
			The momentum conservation~\ref{eq:solutionProcedure-ui} becomes
			\begin{equation}
				\partial_0 \partial_i \variation{u^i} + \frac{4}{3t} \partial_i \variation{u^i} = -v_S^2 \partial_i \partial^i \Delta - v_A^2 \partial_i \partial^i M
			\end{equation}
			and its counterpart along the $ z $-axis remains~\eqref{eq:solutionProcedure-u3}:
			\begin{equation}
				\partial_0 \partial_3 \variation{u^3} + \frac{4}{3t} \partial_3 \variation{u^3} + v_S^2 \partial_3 \partial^3 \Delta = 0 .
			\end{equation}
			We need the Einstein $ 33 $-equation only at $ 0 $-order in the magnetic field, after being multiplied by~$ {v_A^2} $, so equation~\eqref{eq:solutionProcedure-Tord0} in our limit reads
			\begin{equation}
				v_A^2 \partial_i \partial^i \dot{T} + v_A^2 \partial_3 \partial^3 (\partial_i \variation{u^i} + \dot{\Delta}) = 0 .
			\end{equation}
			
			With some algebra it is possible to reduce this system to a single equation. Expanding the spatial part in Fourier, defining the anisotropy parameter~$ {\mu} $ of the solution as
			\begin{equation}
				k_3 k^3 = \mu^2 k^2
			\end{equation}
			and using the constants~\eqref{eq:solutionProcedure-LambdaConstants} we find, after some algebra,
			\begin{equation}\label{eq:solutionProcedure-matterSmallScales-analyticalSolutions-finalEquation}
			\begin{split}
				&9 t^4 \Delta^{(4)} + 60 t^3 \Delta^{(3)}\\
					&\quad + \left[ 76 + 9 \Lambda_S^2 t^{-2\nu} + 18 \Lambda_A^2 \right] t^2 \Delta^{(2)}\\
					&\quad + \left[ 8 + 12 \Lambda_S^2 ( 1 - 3\nu ) t^{-2\nu} + 24 \Lambda_A^2 \right] t \Delta^{(1)}\\
					&\quad + \big[ 6 \Lambda_S^2 \left( - \nu + 6 \nu^2 + 3 \mu^2 \Lambda_A^2 \right) t^{-2\nu}\\
						&\qquad - 12 \mu^2 \Lambda_A^2 \big] \Delta = 0 ,
			\end{split}
			\end{equation}
			where $ {\Delta^{(i)}} $ is the $ {i} $-th derivative of~$ {\Delta} $. This corresponds exactly to equation~(29) of~\cite{bib:lattanzi-2012}, except for a difference in the definition of~$ {v_A^2} $ and so in~$ {\Lambda_A} $.
			
			We believe interesting to analyse separately the two cases of~$ {\nu = 0} $ and~$ {\nu = 1/3} $, instead of studying them together as in~\cite{bib:lattanzi-2012}.
			
		\subsubsection{Post recombination evolution}
			For $ {1100>z>100} $ we have~$ {\nu=0} $. The solution of~\eqref{eq:solutionProcedure-matterSmallScales-analyticalSolutions-finalEquation} is
			\begin{equation}
				\Delta = \Delta_i t^{x_i} ,
			\end{equation}
			where $ {\Delta_i} $ are arbitrary constants and
			\begin{subequations}\label{eq:solutionProcedure-matterSmallScales-analyticalSolutions-postRecombination-exponents}
				\begin{gather}
					x_1 = \left( -1 + \sqrt{\delta_-} \right) / 6 \quad x_2 = \left( -1 - \sqrt{\delta_-} \right) / 6\\
					x_3 = \left( -1 + \sqrt{\delta_+} \right) / 6 \quad x_4 = \left( -1 - \sqrt{\delta_+} \right) / 6\\
					\delta_\pm = \delta_1 \pm 6 \sqrt{ \delta_2 }\\
					\delta_1 = 13 - 18 \Lambda_S^2 - 36 \Lambda_A^2\\
					\delta_2 = \left( -2 + 6 \Lambda_A^2 + 3 \Lambda_S^2 \right)^2 - 24 \mu^2 \Lambda_A^2 \left( -2 + 3  \Lambda_S^2 \right) .
				\end{gather}
			\end{subequations}
			The only possible growing solution is~$ {x_3} $, and the requirement is that it holds one of the conditions
			\begin{subequations}\label{eq:solutionProcedure-matterSmallScales-analyticalSolutions-postRecombinationConditions}
				\begin{gather}
					\mu > 0 \text{ and } \Lambda_S^2 < \frac{2}{3}\\
					\mu = 0 \text{ and } \Lambda_S^2 + 2 \Lambda_A^2 < \frac{2}{3} ;
				\end{gather}
			\end{subequations}
			using~\eqref{eq:solutionProcedure-LambdaConstants} and~\eqref{eq:background-FRWSolution}, making explicit the presence of Newton's constant we get~$ {\rho = 1/6 \pi G t^2} $, conditions~\eqref{eq:solutionProcedure-matterSmallScales-analyticalSolutions-postRecombinationConditions} become
			\begin{subequations}
				\begin{gather}
					\mu > 0 \text{ and } k < k_J = \sqrt{ \frac{ 4 \pi G \rho }{ v_S^2 } }\\
					\mu = 0 \text{ and } k < \sqrt{ \frac{ 4 \pi G \rho }{ v_S^2 + 2 v_A^2 } } < k_J \label{eq:solutionProcedure-postRecombination-magneticJeansLength}.
				\end{gather}
			\end{subequations}
			While the first one is the standard Jeans condition, the second one means that, orthogonally to the background magnetic field, there is a heavier requirement dependent on the strength of the magnetic field: some modes could grow in every direction but the one of the field. The presence of the magnetic field also imposes a slowing down of the growing mode:
			\begin{equation}
				x_3 \leq  \left. x_3 \right|_{\Lambda_A=0} = \frac{1}{6} \left( -1+\sqrt{25-36\Lambda_S^2} \right) ,
			\end{equation}
			where the equal sign holds only in absence of a magnetic field, that is only if~$ {\Lambda_A = 0} $.
		
		\subsubsection{Late times evolution}
			This is exactly the case analysed in~\cite{bib:lattanzi-2012}. For~$ {z<100} $ we have~$ {\nu>0} $ and the solution of~\eqref{eq:solutionProcedure-matterSmallScales-analyticalSolutions-finalEquation} is
			\begin{equation}\label{eq:solutionProcedure-matterSmallScales-analyticalSolutions-lateTimesSolution}
				\Delta = \Delta_i t^{x_i} {}_2 F_3 \left[ 
					\begin{matrix}
						a_{i1},a_{i2}\\
						b_{i1},b_{i2},b_{i3}
					\end{matrix} ;
					- \frac{ \Lambda_S^2 t^{-2\nu} }{ 4\nu^2 } \right] ,
			\end{equation}
			where $ {\Delta_i} $ are arbitrary constants, $ {}_2 F_3 $ is a generalized hypergeometric function with constant coefficients~$ {a_{ij},\ b_{ij}} $ depending only on the constants~$ {\nu,\ \Lambda_S,\ \Lambda_A} $ (see app.~\ref{app:lateTimesSolution} for the explicit value of the coefficients) and
			\begin{subequations}\label{eq:solutionProcedure-matterSmallScales-analyticalSolutions-lateTimesSolution-exponents}
				\begin{gather}
					x_1 = \left( -1 + \sqrt{\delta_-} \right)/6 \quad x_2 = \left( -1 - \sqrt{\delta_-} \right)/6\\
					x_3 = \left( -1 + \sqrt{\delta_+} \right)/6 \quad x_4 = \left( -1 - \sqrt{\delta_+} \right)/6\\
					\label{eq:solutionProcedure-matterSmallScales-analyticalSolutions-lateTimesSolution-exponents-delta}\delta_\pm = 13 - 36 \Lambda_A^2 \pm 12 \sqrt{ \left( 1 - 3 \Lambda_A^2 \right)^2 + 12 \mu^2 \Lambda_A^2 } .
				\end{gather}
			\end{subequations}
			
			The solutions can grow only if the argument of the hypergeometric functions is small, i.e.\ if
			\begin{equation}\label{eq:solutionProcedure-matterSmallScales-analyticalSolutions-lateTimesSolution-jeansGrowingCondition}
				\Lambda_S^2 / 4 \nu^2 t^{2\nu} \ll 1 :
			\end{equation}
			this way we have
			\begin{equation}\label{eq:solutionProcedure-matterSmallScales-analyticalSolutions-lateTimesSolution-jeansGrowingModes}
				\Delta = \Delta_i t^{x_i} \left( 1 + \order{ \frac{ \Lambda_S^2 t^{-2\nu} }{ 4\nu^2 } } \right) .
			\end{equation}
			Condition~\eqref{eq:solutionProcedure-matterSmallScales-analyticalSolutions-lateTimesSolution-jeansGrowingCondition} is the standard Jeans condition~\cite{bib:weinberg-gravitationAndCosmology}: using~\eqref{eq:solutionProcedure-LambdaConstants} and~\eqref{eq:background-FRWSolution}, eq.~\eqref{eq:solutionProcedure-matterSmallScales-analyticalSolutions-lateTimesSolution-jeansGrowingCondition}~translates in~\cite{bib:lattanzi-2012}
			\begin{equation}
				k \ll k_J = \sqrt{ \frac{ 24 \nu^2 \pi G \rho }{ v_S^2 } }  .
			\end{equation}
			
			The only solution in~\eqref{eq:solutionProcedure-matterSmallScales-analyticalSolutions-lateTimesSolution-jeansGrowingModes} that can grow is~$ 3 $: $ x_3 > 0 $ only if it holds one of
			\begin{subequations}
				\begin{gather}
					0 < \mu \leq 1\\
					\mu = 0 \text{ and } \Lambda_A^2 < \frac{1}{3} .
				\end{gather}
			\end{subequations}
			The first one means that, in any direction but orthogonal to the background magnetic field, the only necessary condition is the standard one. The second one is an additional condition that must hold for perturbations propagating orthogonally to the background magnetic field, and using~\eqref{eq:solutionProcedure-LambdaConstants} and~\eqref{eq:background-FRWSolution} it reads~\cite{bib:lattanzi-2012}
			\begin{equation}\label{eq:solutionProcedure-lateTimes-magneticJeansLength}
				k < k_A = \sqrt{ \frac{ 2 \pi G \rho }{ v_A^2 } } .
			\end{equation}
			The presence of this new condition makes possible the existence of Jeans unstable modes, that orthogonally to the background magnetic field are stabilized by the magnetic pressure if~$ {k_A < k_J} $ and~$ {k_A < k < k_J} $~\cite{bib:lattanzi-2012}.
			
			Studying the growing rate of this solution with more care, we see that $ {x_3} $~satisfies
			\begin{subequations}
				\begin{gather}
					\mu = 1 \quad \implies \quad x_3 = \left. x_3 \right|_{\Lambda_A = 0} = \frac{2}{3}\\
					\mu \neq 0 \quad \implies \quad x_3 < \left. x_3 \right|_{\Lambda_A = 0} :
				\end{gather}
			\end{subequations}
			orthogonally to the background magnetic field the growing rate is unchanged, while in other directions it is slowed down, depending on the field strength.
			
	\subsection{Full relativistic case}
		If we put $ v_S^2 = 0 $ we recover the exact relativistic solution. As we can see from the previous solutions, the growing condition is
		\begin{subequations}
			\begin{gather}
				\mu > 0\\
				\mu = 0 \text{ and } k < k_A = \sqrt{ \frac{ 2 \pi G \rho }{ v_A^2 } }.
			\end{gather}
		\end{subequations}
		Moreover, the solution is
		\begin{equation}
		\label{eq:solutionProcedure-fullRelativisticCase-solution}
			\Delta = \Delta_i t^{x_i}
		\end{equation}
		with $ x_i $ given by~\eqref{eq:solutionProcedure-matterSmallScales-analyticalSolutions-postRecombination-exponents} with $ {\Lambda_S = 0} $, or equivalently by~\eqref{eq:solutionProcedure-matterSmallScales-analyticalSolutions-lateTimesSolution-exponents}.
		
		If we compare our result with~\cite{bib:barrow-2007}, we identify the\linebreak anisotropic behaviour and we obtain the correct Newtonian limit of~\cite{bib:lattanzi-2012}. However, our solutions are different and we are unable to explain such discrepancy: we can argue they may have found some sort of average effect, however this is not clear, given the strong anisotropy of the model: the magnetic Jeans wavenumber is present only in one direction, the one with~$ {\mu = 0} $.
		
		In case~$ {\Lambda_A^2 \ll 1} $ we have
		\begin{subequations}
		\begin{gather}
			x_\doubleCoefficient{1}{2} = \frac{1}{6} \left( -1 \pm \sqrt{1 - 72 \mu^2 \Lambda_A^2} \right)\\
			x_\doubleCoefficient{3}{4} = \frac{1}{6} \left( -1 \pm \sqrt{25 - 72 (1 - \mu^2) \Lambda_A^2} \right)
		\end{gather}
		\end{subequations}
		and setting~$ {\mu^2 = 1/3} $ the solutions $ {x_3} $ and $ {x_4} $ recover eq.~(31) of~\cite{bib:vasileiou-2015} and eq.~(31) of~\cite{bib:tseneklidou-2018}, so our small scales solution of sec.~\ref{ssec:solutionProcedure-matterSmallScales} is a generalization of their work, while including a nonvanishing sound speed and pressure.
		
\section{Numerical integration\label{sec:numericalIntegration}}
	To better show our results, we numerically integrated the system~\eqref{eq:perturbedEquations-fluidEnergyConservation}--\eqref{eq:perturbedEquations-momentumConservationDerivative3}, using estimates from~\cite{bib:planck-2018} to set the numerical values for the background functions. We followed the same procedure of~\cite{bib:lattanzi-2012} to determine the initial conditions: we started the integration from a very early time and we verified that the initial perturbations were outside the Hubble horizon and we used the large scale solution to match the initial conditions to the growing mode; in our case such conditions come from eq.~\eqref{eq:solutionProcedure-largeScalesSolutions}.
	
	We assumed to perturb only the baryon component\linebreak of the universe, while leaving the CDM component\linebreak unperturbed; a rigorous treatment should rely on a multi--fluid model, but we ague that we can still extract meaningful information within our approximation. Practically speaking, this assumption means that every quantity present in our equations at perturbative level must be replaced by its baryonic component, while the background model still depends on CDM. Our equations are still correct, because the background interaction is only due to energy density, while at perturbative level every dependence on CDM disappears, except from background quantities.
	
	We choosed to study the same scales of~\cite{bib:lattanzi-2012},\linebreak i.e.\ $ {k \approx ( 17, 1.7, 0.37 )\ \si{\mega\parsec^{-1}}} $ normalized at present\linebreak time, corresponding to baryonic masses of\linebreak $ {M \approx ( \num{1.5e8}, \num{1.5e11}, \num{1.5e13} )\ \mathrm{M}_\odot} $ and roughly\linebreak equivalent respectively to a dwarf galaxy, a galaxy and a galaxy cluster. The results of the numerical integration are shown in figure~\ref{fig:numericalIntegration}.
	\afterpage{
		\begin{figure}[H]
			\centering
			\subfloat[Perturbations at dwarf galaxy scale: $ {k \simeq \SI{17}{\mega\parsec^{-1}}} $, $ {M \approx \num{1.5e8}\ \mathrm{M}_\odot} $.\label{fig:numericalIntegration-17}]{
				\includegraphics[width=\columnwidth,keepaspectratio]{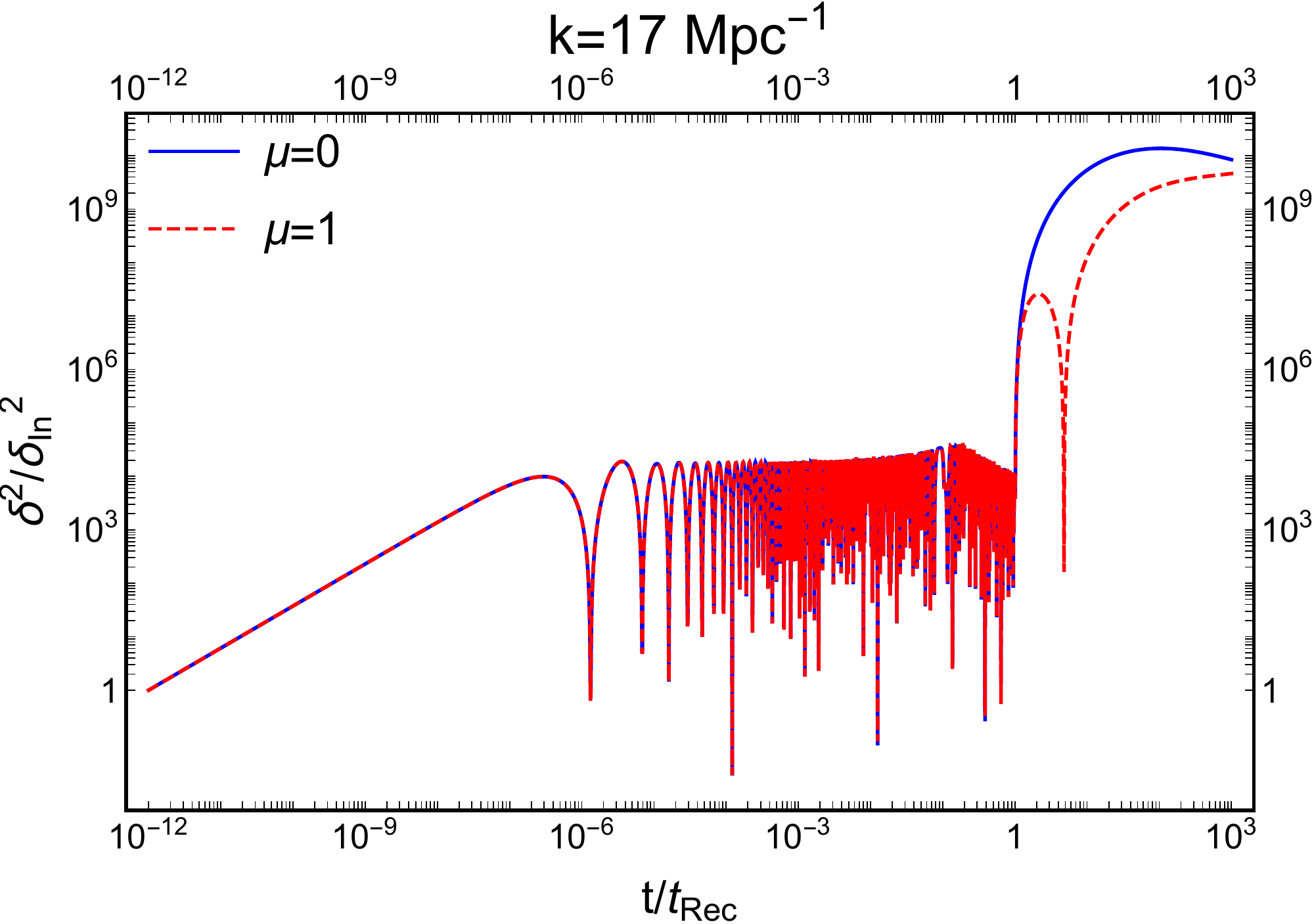}
			}\\
			\subfloat[Perturbations at galactic scale: $ {k \simeq \SI{1.7}{\mega\parsec^{-1}}} $, $ {M \approx \num{1.5e11}\ \mathrm{M}_\odot} $.\label{fig:numericalIntegration-1.7}]{
				\includegraphics[width=\columnwidth,keepaspectratio]{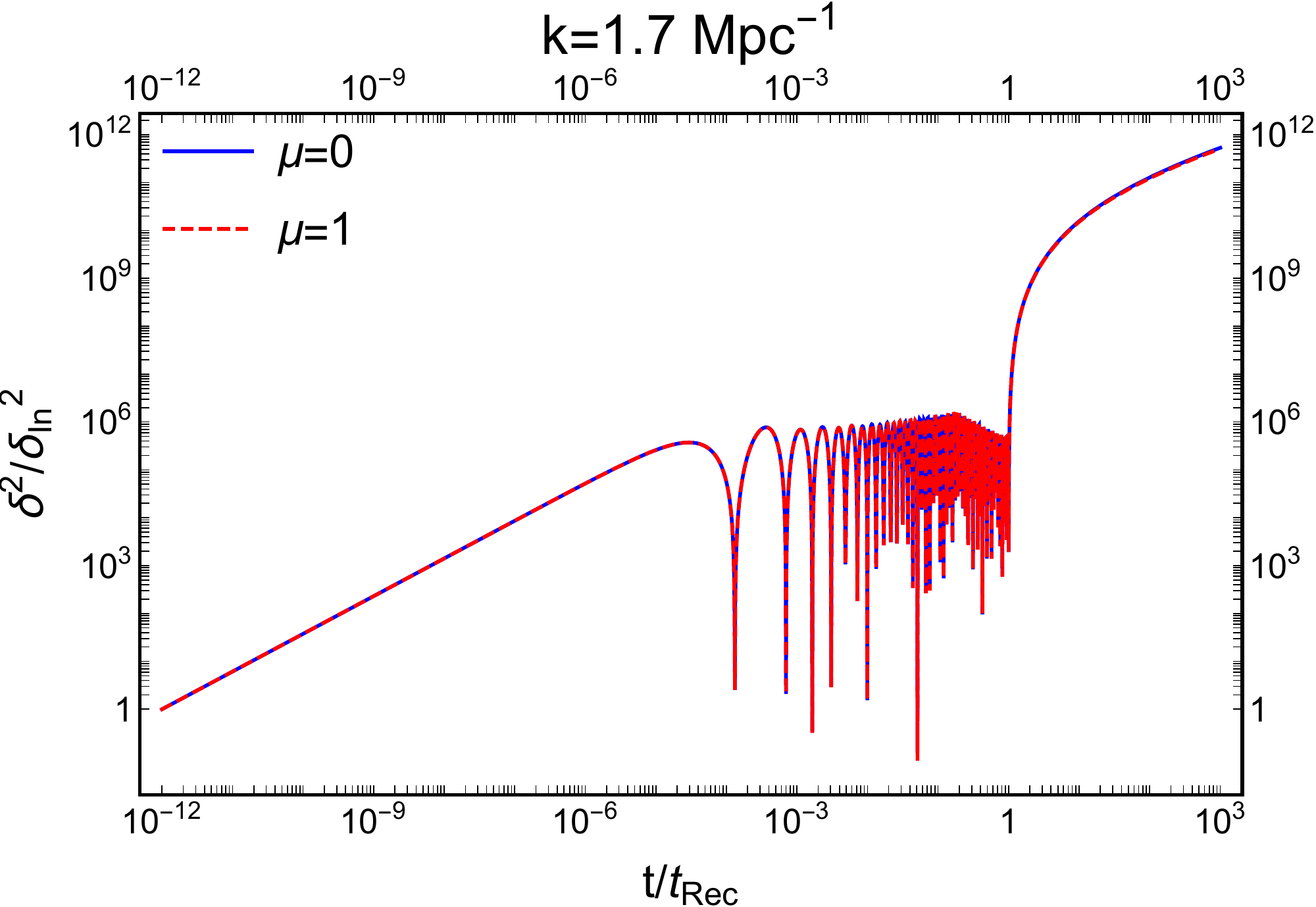}
			}\\
			\subfloat[Perturbations at galaxy cluster scale: $ {k \simeq \SI{0.37}{\mega\parsec^{-1}}} $, $ {M \approx \num{1.5e13}\ \mathrm{M}_\odot} $.]{
				\includegraphics[width=\columnwidth,keepaspectratio]{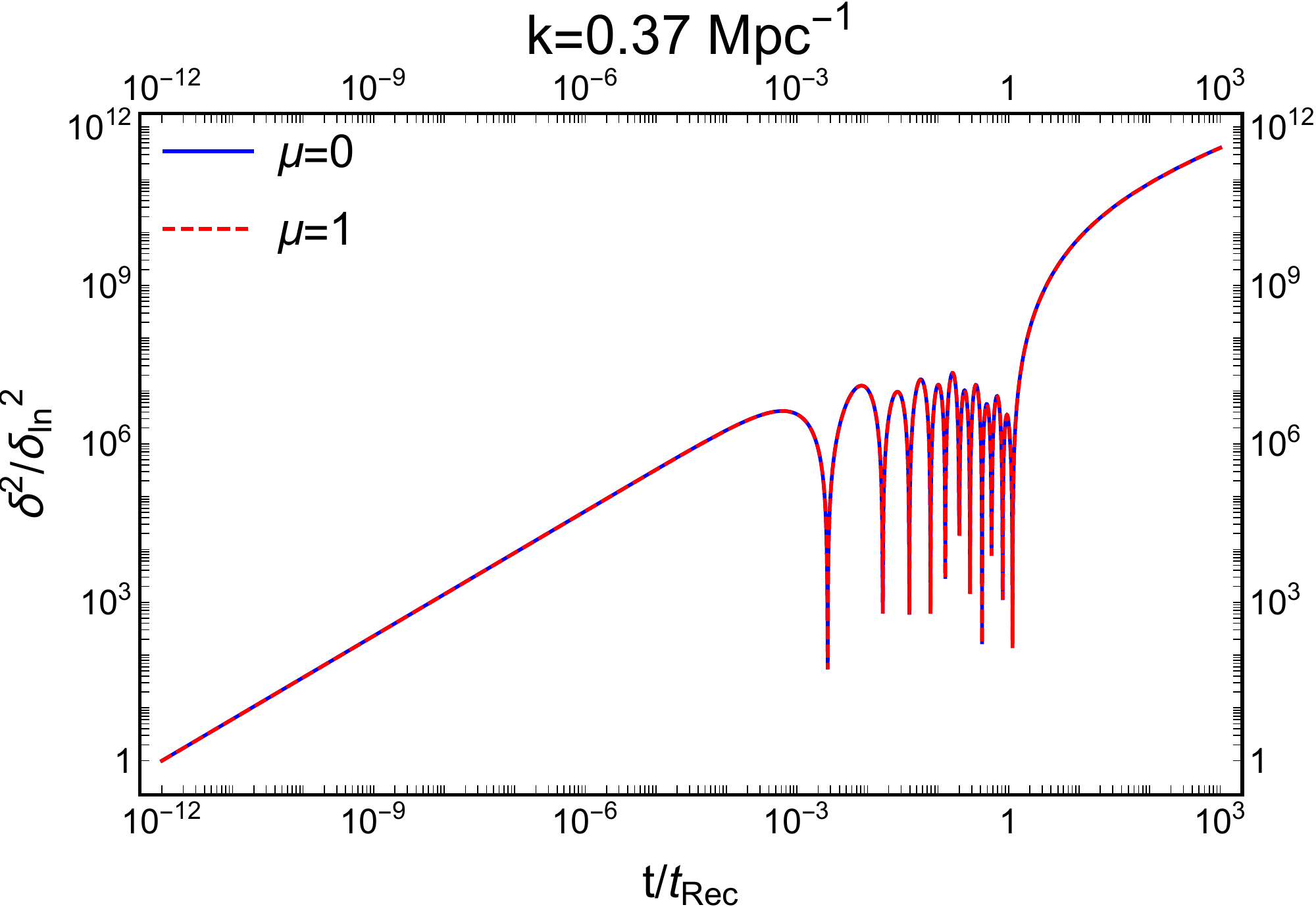}
			}
			\caption{Density perturbations evolution in time, relative to their initial value. While some anisotropy is present in~\protect\subref{fig:numericalIntegration-17} because of the magnetic Jeans length (see sec.~\ref{sec:numericalIntegration} and~\cite{bib:lattanzi-2012}), most of the anisotropic effects of~\cite{bib:lattanzi-2012} here are suppressed because of thermal pressure in the radiation dominated era.\label{fig:numericalIntegration}}
		\end{figure}
	}
	
	Our results must be compared to the ones of~\cite{bib:lattanzi-2012}. Until equivalence ($ {z \approx 3400} $) we are in radiation dominated universe and the comparison is obvious: our solutions grow, while theirs decay; this is because in~\cite{bib:lattanzi-2012} the authors always consider matter dominated universe.
	
	After equivalence, in both cases we are subject to a decaying period, followed by a new growth after recombination, but in our case this happens for a shorter time; most of the anisotropic effects comes in this era, because before equivalence the thermal pressure is much stronger than the magnetic one and most of the anisotropy is suppressed, so they are less relevant in our simulations. This is clear in fig.~\ref{fig:numericalIntegration-1.7}, where we see almost no anisotropy. As a further confirmation, it can be shown that $ {\Delta(z \approx 10)/ \Delta( z \approx 1100 )} $\linebreak has the same value in both the analysed cases, so\linebreak the main anisotropic contribution comes from the\linebreak region~$ {3400 \lesssim z \lesssim 1100} $.
	
	After recombination we have a behaviour similar to~\cite{bib:lattanzi-2012}, because here we are at scales were the Newtonian approximation is correct. The apparent discrepancy in the oscillating behaviour of fig.~\ref{fig:numericalIntegration-17} is mainly due to the (small) difference in the numerical values of the background functions, because the oscillating behaviour is very sensible to such numbers; however, the qualitative evolution is the same, with the $ {\mu = 0} $ case beginning to decay because of the magnetic Jeans length~\cite{bib:lattanzi-2012} (eq.~\eqref{eq:solutionProcedure-postRecombination-magneticJeansLength} and~\eqref{eq:solutionProcedure-lateTimes-magneticJeansLength}). In this region our solution has a slightly faster growth than~\cite{bib:lattanzi-2012}, we argue this may be caused by some residual relativistic effects, but it has to be investigated with more care.
	
	Moreover, in the last region we should be outside of the linear regime, so we would need a full nonlinear treatment.

\section{Conclusion\label{sec:conclusion}}
	We developed above a self-consistent scheme for the analysis of cosmological perturbations in the presence of a magnetic field. We set up in the synchronous gauge a dynamical scheme which accounts for the effects induced by the magnetic field both on the background and the first order formulation. To this end, we considered a Bianchi~I model, whose anisotropy with respect to the flat Robertson-Walker geometry is due to the privileged direction defined by the magnetic field.
	
	We first solve in detail the equations describing the\linebreak anisotropic background and then we analyse the perturbation dynamics, having awareness of the gauge contribution analytical form.
	
	We amended for the previous analysis in~\cite{bib:barrow-2007} in the case of a super-horizon wavelength of the perturbation. In particular, our solution has a clearer comparison with the non magnetic one. We recovered the slowing-down of the growing mode caused by the magnetic pressure, and so of order~$ \order{v_A^2} \ll 1 $. This effect has long been known in FRW models and has been analysed in Bianchi~I models with particular anisotropies by~\cite{bib:tsagas-2000}, while we worked always relating the background anisotropy to the magnetic field without additional assumptions.
	
	We refined the results of~\cite{bib:barrow-2007} for the sub-horizon wavelength of the perturbations, showing that an anisotropic treatment is required. We also generalised the results of~\cite{bib:vasileiou-2015} and~\cite{bib:tseneklidou-2018}, while including a nonvanishing sound speed and considering the anisotropic case.
	
	We finally enforced the Newtonian limit obtained in~\cite{bib:lattanzi-2012}, completing it with the relativistic analysis, also facing a numerical treatment. We showed that the relativistic regime limits the anisotropy induced by the magnetic field.
	
	Overall, despite the assumption of a Bianchi~I\linebreak background, most of our solutions reproduce those obtained on an FRW background. At a closer look, the Bianchi~I anisotropy enters the system via the $ S $~function defined in~\eqref{eq:background-definition-HS}. At small scales the relevant terms are the ones with~$ {k^2} $, and none of those are related to such anisotropy. However, when the condition $ H^2 \ll k^2 $ does not hold, such terms become important; unfortunately, in this case the system would be much more complicated that the one of sec.~\ref{ssec:solutionProcedure-matterSmallScales}. On the other hand, at large scales the background anisotropy survives, and we argue that it is mainly related to the perturbed fluid velocity. In particular, it can be shown that the solutions proportional to~$ t^{1/2} $ in~\eqref{eq:solutionProcedure-largeScalesSolutionsComplete} are related to~$ {\variation{u^i}} $, and more precisely in $ {\Delta_2 t^{1/2+4 v_{A0}^2}} $ we have both $ {k_i \variation{u^i} \neq 0} $ and $ {k_3 \variation{u^3} \neq 0} $, while in $ \Delta_1 t^{1/2-2 v_{A0}^2} $ it holds $ {k_3 \variation{u^3} = 0} $; the solutions $ {\Delta_\textrm{grow} t^{1-2 v_{A0}^2}} $ and $ {\Delta_\textrm{gauge}/t} $, on the other hand, both have~$ {k_i \variation{u^i} = k_3 \variation{u^3} = 0} $.
	
	We stress that, in order to solve the equations, we assumed a small magnetic field and so all the effects we studied are related to~$ {v_A^2 \ll 1} $, and they become relevant only at small scales, due to the large wavenumber~$ {k^2 \gg H^2 } $ and to the also small sound speed~$ v_S^2 $. This is clear by looking at fig.~\ref{fig:velocities}.

\appendix

\section{Magnetic field at perturbative level\label{app:magneticField}}
	In literature there are different definitions of the magnetic field at a perturbative level, but it is easy to recognize that not all of them satisfy the required properties. After a careful analysis we concluded that the correct one, at least with respect to the physical phenomenon we study here, it the one of~\cite{bib:barrow-2007} made through the 3+1 formalism. This way, the magnetic field is defined as the spatial projected part of the Faraday tensor~$ F_{\mu\nu} $, while the electric field as the temporal one
	\begin{subequations}
	\begin{gather}
		E_\mu = F_{\mu\nu} u^\nu\\
		B_\mu = \frac{1}{2} \epsilon_{\mu\nu\rho} F^{\nu\rho} = \frac{1}{2} \eta_{\mu\nu\rho\sigma} F^{\nu\rho} u^\sigma
	\end{gather}
	\end{subequations}
	and we have
	\begin{equation}
		B_\mu u^\mu = 0
	\end{equation}
	at all orders.
		
	There are two important reasons for this requirement. The first one is that the electromagnetic field is decomposed in electric and magnetic components by the observer and we are interested in its interaction with the cosmological fluid, so the natural observer is the fluid itself. Beside that, we force a vanishing electric field $ E_\mu = 0 $ through the assumption of infinite conductivity of the medium, thus we work in the limit of ideal MHD. To do this we need these fields to be defined with respect to the fluid. Using this definition there are no induced fields, reflecting the fact that the covariant form of Maxwell’s formulae and of the electric and magnetic field definitions already incorporates the effects of relative motion~\cite{bib:barrow-2007}.
	
	The second reason is that with different definitions we would have a nonvanishing trace for the perturbed magnetic stress energy tensor, while this way all goes well and it is traceless. This is easy to check using the definition of perturbations from section~\ref{sec:perturbedEquations}.
	
\section{Gauge behaviour in late times\label{app:lateTimesGauge}}
	We will analyse here the FRW case, to clarify the meaning of~$ \variation{\rho} $ becoming gauge invariant for late times. Following~\cite{bib:weinberg-gravitationAndCosmology} and using the Newtonian approximation we see that the solutions after recombination are
	\begin{equation}\label{key}
		\delta_\pm \propto t^{-1/6} J_{\mp\frac{5}{6\nu}} \left( \frac{\Lambda t^{-\nu}}{\nu} \right) ,
	\end{equation}
	where $ \gamma = \nu + 4/3 > 4/3 $ is the heat ratio of the fluid (after recombination $ \gamma \simeq 5/3 $), $ \delta = \variation{\rho} / \rho $,
	\begin{equation}
		\Lambda = t^{2\gamma - 2/3} v_S^2 k^2
	\end{equation}
	is a constant, $ v_S^2 $ is the squared sound speed and $ k $ the\linebreak wavenumber. The functions~$ J_a (z) $ are the Bessel functions: when their argument is large they oscillate, but when the argument is small they behave like
	\begin{equation}
		\delta_\pm \propto t^{(-1 \pm 5) / 6} .
	\end{equation}
	The growing mode is the physical solution we are looking for, while the other one decays to zero.
	
	We cannot speak of gauge modes in Newtonian theory, but the decaying mode corresponds exactly to the relativistic gauge mode, and as expected it decays in time with respect to the growing one. This means that, for large times, gauge modes naturally decay to zero and we can neglect them as long as we are looking only for the growing ones.
	
	It should be noted that in our calculations we are in the same situation: we cannot have a relativistic sound speed different from $ w $ in a single fluid model, but we make this approximation in section~\ref{sec:solutionProcedure} because from a physical point of view we need a nonvanishing sound speed. This way we ``break'' the gauge invariance, but the gauge modes manifest themselves in one of the decaying solutions. We are only looking for growing modes, so we can safely neglect them.
	
\section{Late times solution coefficients\label{app:lateTimesSolution}}
	We report here the values of the coefficients of the hypergeometric function appearing in~\eqref{eq:solutionProcedure-matterSmallScales-analyticalSolutions-lateTimesSolution}, using $ {\delta_\pm} $ defined in eq.~\eqref{eq:solutionProcedure-matterSmallScales-analyticalSolutions-lateTimesSolution-exponents-delta}:
	\begin{subequations}
		\begin{gather}
			a_{\doubleCoefficient{1}{2} 1} = 1 \mp \sqrt{\delta_-} / 12\nu - \sqrt{1 - 72 \mu^2 \Lambda_A^2} / 12\nu\\
			a_{\doubleCoefficient{1}{2} 2} = 1 \mp \sqrt{\delta_-} / 12\nu + \sqrt{1 - 72 \mu^2 \Lambda_A^2} / 12\nu\\
			a_{\doubleCoefficient{3}{4} 1} = 1 \mp \sqrt{\delta_+} / 12\nu - \sqrt{1 - 72 \mu^2 \Lambda_A^2} / 12\nu\\
			a_{\doubleCoefficient{3}{4} 2} = 1 \mp \sqrt{\delta_+} / 12\nu + \sqrt{1 - 72 \mu^2 \Lambda_A^2} / 12\nu\\
			b_{\doubleCoefficient{1}{2} 1} = 1 \mp \sqrt{\delta_-} / 6\nu\\
			b_{\doubleCoefficient{1}{2} 2} = 1 \mp \sqrt{\delta_-} / 12\nu - \sqrt{\delta_+} / 12\nu\\
			b_{\doubleCoefficient{1}{2} 3} = 1 \mp \sqrt{\delta_-} / 12\nu + \sqrt{\delta_+} / 12\nu\\
			b_{\doubleCoefficient{3}{4} 1} = 1 \mp \sqrt{\delta_+} / 6\nu\\
			b_{\doubleCoefficient{3}{4} 2} = 1 \mp \sqrt{\delta_+} / 12\nu - \sqrt{\delta_-} / 12\nu\\
			b_{\doubleCoefficient{3}{4} 3} = 1 \mp \sqrt{\delta_+} / 12\nu + \sqrt{\delta_-} / 12\nu .
		\end{gather}
	\end{subequations}

\bibliographystyle{spphys} 
\bibliography{ms}

\begin{thebibliography}{10}
\providecommand{\url}[1]{{#1}}
\providecommand{\urlprefix}{URL }
\expandafter\ifx\csname urlstyle\endcsname\relax
  \providecommand{\doi}[1]{DOI \discretionary{}{}{}#1}\else
  \providecommand{\doi}{DOI \discretionary{}{}{}\begingroup
  \urlstyle{rm}\Url}\fi

\bibitem{bib:armendariz-picon-2014}
C.~Armendariz-Picon, J.T. Neelakanta, Journal of Cosmology and Astroparticle
  Physics \textbf{3}, 049 (2014).
\newblock \doi{10.1088/1475-7516/2014/03/049}

\bibitem{bib:gheller-2016}
C.~Gheller, F.~Vazza, M.~Brüggen, M.~Alpaslan, B.W. Holwerda, A.~Hopkins,
  J.~Liske, Mon.\ Not.\ Roy.\ Astron.\ Soc. \textbf{462}(1), 448 (2016).
\newblock \doi{10.1093/mnras/stw1595}

\bibitem{bib:dickau-2009}
J.J. Dickau, Chaos, Solitons \& Fractals \textbf{41}(4), 2103  (2009).
\newblock \doi{10.1016/j.chaos.2008.07.056}

\bibitem{bib:grujic-2009}
P.~{Grujic}, V.~{Pankovic}, ArXiv e-prints \textbf{physics.gen-ph}, 0907.2127
  (2009)

\bibitem{bib:banerjee-2004}
R.~Banerjee, K.~Jedamzik, Phys.\ Rev.\ D \textbf{70}, 123003 (2004).
\newblock \doi{10.1103/PhysRevD.70.123003}

\bibitem{bib:lattanzi-2012}
M.~Lattanzi, N.~Carlevaro, G.~Montani, Phys.\ Lett.\ B \textbf{718}(2), 255
  (2012).
\newblock \doi{10.1016/j.physletb.2012.10.067}

\bibitem{bib:giovannini-2004}
M.~{Giovannini}, International Journal of Modern Physics D \textbf{13}(03), 391
  (2004).
\newblock \doi{10.1142/S0218271804004530}

\bibitem{bib:kosowsky-1996}
A.~{Kosowsky}, A.~{Loeb}, Astrophys.\ J. \textbf{469}, 1 (1996).
\newblock \doi{10.1086/177751}

\bibitem{bib:barrow-1997-constraints}
J.D. {Barrow}, P.G. {Ferreira}, J.~{Silk}, Phys.\ Rev.\ Lett. \textbf{78}, 3610
  (1997).
\newblock \doi{10.1103/PhysRevLett.78.3610}

\bibitem{bib:barrow-1997}
J.D. {Barrow}, Phys.\ Rev.\ D \textbf{55}, 7451 (1997).
\newblock \doi{10.1103/PhysRevD.55.7451}

\bibitem{bib:komatsu-2011}
E.~{Komatsu}, K.M. {Smith}, J.~{Dunkley}, C.L. {Bennett}, B.~{Gold},
  G.~{Hinshaw}, N.~{Jarosik}, D.~{Larson}, M.R. {Nolta}, L.~{Page}, D.N.
  {Spergel}, M.~{Halpern}, R.S. {Hill}, A.~{Kogut}, M.~{Limon}, S.S. {Meyer},
  N.~{Odegard}, G.S. {Tucker}, J.L. {Weiland}, E.~{Wollack}, E.L. {Wright}, The
  Astrophysical Journal Supplement Series \textbf{192}(2), 18 (2011).
\newblock \doi{10.1088/0067-0049/192/2/18}

\bibitem{bib:paoletti-2011}
D.~{Paoletti}, F.~{Finelli}, Phys.\ Rev.\ D \textbf{83}(12), 123533 (2011).
\newblock \doi{10.1103/PhysRevD.83.123533}

\bibitem{bib:pogosian-2014}
L.~{Pogosian}, Journal of Physics: Conference Series \textbf{496}(1), 012025
  (2014).
\newblock \doi{10.1088/1742-6596/496/1/012025}

\bibitem{bib:planck-2015}
{Planck Collaboration}, et~al., Astronomy \& Astrophysics \textbf{594}, A19
  (2016).
\newblock \doi{10.1051/0004-6361/201525821}

\bibitem{bib:montani-2017}
G.~{Montani}, G.~{Palermo}, N.~{Carlevaro}, ArXiv e-prints  (2017)

\bibitem{bib:weinberg-gravitationAndCosmology}
S.~{Weinberg}, \emph{Gravitation and Cosmology: Principles and Applications of
  the General Theory of Relativity} (John Wiley \& Sons, 1972)

\bibitem{bib:kolb-turner}
E.~Kolb, M.~Turner, \emph{The Early Universe}.
\newblock Frontiers in Physics (Westview Press, 1994)

\bibitem{bib:weinberg-cosmology}
S.~Weinberg, \emph{Cosmology} (Oxford University Press, 2008)

\bibitem{bib:barrow-2007}
J.D. {Barrow}, R.~{Maartens}, C.G. {Tsagas}, Physics Reports \textbf{449}(6),
  131  (2007).
\newblock \doi{10.1016/j.physrep.2007.04.006}

\bibitem{bib:vasileiou-2015}
H.~Vasileiou, C.G. Tsagas, Monthly Notices of the Royal Astronomical Society
  \textbf{455}(3), 2500 (2015).
\newblock \doi{10.1093/mnras/stv2418}

\bibitem{bib:tseneklidou-2018}
D.~{Tseneklidou}, C.G. {Tsagas}, J.D. {Barrow}, Classical and Quantum Gravity
  \textbf{35}(12), 124001 (2018).
\newblock \doi{10.1088/1361-6382/aac07f}

\bibitem{bib:tsagas-2000}
C.G. {Tsagas}, R.~{Maartens}, Classical and Quantum Gravity \textbf{17}(11),
  2215 (2000).
\newblock \doi{10.1088/0264-9381/17/11/305}

\bibitem{bib:belinskii-1982}
V.A. Belinskii, I.M. Khalatnikov, E.M. Lifshitz, Advances in Physics
  \textbf{31}(6), 639 (1982).
\newblock \doi{10.1080/00018738200101428}

\bibitem{bib:landau-classicalFields}
L.D. Landau, E.M. Lifshitz, \emph{The Classical Theory of Fields}, \emph{Course
  of Theoretical Physics}, vol.~2, 4th edn. (Elsevier Science, 2013)

\bibitem{bib:zeldovichNovikov-relativisticAstrophysics2TheStructureAndEvolutionOfTheUniverse}
I.B. {Ze'ldovich}, I.D. {Novikov}, \emph{Relativistic Astrophysics, 2: The
  Structure and Evolution of the Universe}, vol.~2, revised and enlarged
  edition edn. (The University of Chicago Press, 1983)

\bibitem{bib:montani-primordialCosmology}
G.~{Montani}, M.V. {Battisti}, R.~{Benini}, G.~{Imponente}, \emph{Primordial
  Cosmology} (World Scientific, 2011).
\newblock \doi{10.1142/7235}

\bibitem{bib:ryanShepley-homogeneousRelativisticCosmologies}
M.P. {Ryan}, L.C. {Shepley}, \emph{Homogeneous relativistic cosmologies}
  (Princeton University Press, 1975)

\bibitem{bib:kirillov-2002}
A.A. Kirillov, G.~Montani, Phys.\ Rev.\ D \textbf{66}, 064010 (2002).
\newblock \doi{10.1103/PhysRevD.66.064010}

\bibitem{bib:leblanc-1997}
V.G. LeBlanc, Classical and Quantum Gravity \textbf{14}(8), 2281 (1997).
\newblock \doi{10.1088/0264-9381/14/8/025}

\bibitem{bib:thorne-1967}
K.S. {Thorne}, Astrophysical Journal \textbf{148}, 51 (1967).
\newblock \doi{10.1086/149127}

\bibitem{bib:jacobs-1969}
K.C. {Jacobs}, Astrophysical Journal \textbf{155}, 379 (1969).
\newblock \doi{10.1086/149875}

\bibitem{bib:king-2007}
E.J. {King}, P.~{Coles}, Classical and Quantum Gravity \textbf{24}(8), 2061
  (2007).
\newblock \doi{10.1088/0264-9381/24/8/008}

\bibitem{bib:stephaniKramer-exactSolutionsOfEinsteinsFieldEquations}
H.~Stephani, D.~Kramer, M.~MacCallum, C.~Hoenselaers, E.~Herlt, \emph{Exact
  solutions of Einstein's field equations}, 2nd edn. (Cambridge University
  Press, 2003)

\bibitem{bib:ehlers-1961}
J.~{Ehlers}, General Relativity and Gravitation \textbf{25}(12), 1225 (1993).
\newblock \doi{10.1007/BF00759031}

\bibitem{bib:ellis-1971}
G.F. {Ellis}, General Relativity and Gravitation \textbf{41}(3), 581 (2009).
\newblock \doi{10.1007/s10714-009-0760-7}

\bibitem{bib:ellis-1973}
G.F. {Ellis}, in \emph{Carg{\`e}se lectures in Physics}, vol.~VI, ed. by
  E.~Schatzmann (Gordon and Breach, 1973), vol.~VI, pp. 1--60

\bibitem{bib:ellis-1998}
G.F.R. {Ellis}, H.~{van Elst}, in \emph{Theoretical and Observational
  Cosmology}, \emph{NATO Advanced Science Institutes (ASI) Series C}, vol. 541,
  ed. by M.~{Lachi{\`e}ze-Rey} (1999), \emph{NATO Advanced Science Institutes
  (ASI) Series C}, vol. 541, pp. 1--116

\bibitem{bib:tsagas-1997}
C.G. {Tsagas}, J.D. {Barrow}, Classical and Quantum Gravity \textbf{14}(9),
  2539 (1997).
\newblock \doi{10.1088/0264-9381/14/9/011}

\bibitem{bib:tsagas-2005}
C.G. {Tsagas}, Classical and Quantum Gravity \textbf{22}(2), 393 (2005).
\newblock \doi{10.1088/0264-9381/22/2/011}

\bibitem{bib:tsagas-2008}
C.G. {Tsagas}, A.~{Challinor}, R.~{Maartens}, Physics Reports
  \textbf{465}(2–3), 61  (2008).
\newblock \doi{10.1016/j.physrep.2008.03.003}

\bibitem{bib:doroshkevich-1965}
A.G. {Doroshkevich}, Astrophysics \textbf{1}(3), 138 (1965).
\newblock \doi{10.1007/BF01041937}

\bibitem{bib:noh-1995}
H.~Noh, J.~Hwang, Phys.\ Rev.\ D \textbf{52}, 1970 (1995).
\newblock \doi{10.1103/PhysRevD.52.1970}

\bibitem{bib:bonometto-1974}
S.A. {Bonometto}, L.~{Danese}, F.~{Lucchin}, Astronomy and Astrophysics
  \textbf{35}, 267 (1974)

\bibitem{bib:planck-2018}
{Planck Collaboration}, et~al., arXiv e-prints arXiv:1807.06209 (2018)

\end{thebibliography}

\end{document}